\documentstyle[12pt,psfig]{article}
\topmargin - 1.0 true cm
\def\sss{\scriptscriptstyle}
\def\barp{{\raise.35ex\hbox{${\sss (}$}}---{\raise.35ex\hbox{${\sss )}$}}}
\def\bdbarp{\hbox{$B_d$\kern-1.4em\raise1.4ex\hbox{\barp}}}
\def\bsbarp{\hbox{$B_s$\kern-1.4em\raise1.4ex\hbox{\barp}}}
\newcommand{\xd}{x_d}
\newcommand{\xs}{x_s}
\newcommand{\bd}{B_d^0}
\newcommand{\bdb}{\overline{B_d^0}}
\newcommand{\bs}{B_s^0}
\newcommand{\bsb}{\overline{B_s^0}}
\newcommand{\bu}{B_u^\pm}
\newcommand{\beq}{\begin{equation}}
\newcommand{\eeq}{\end{equation}}
\newcommand{\absvcb}{\vert V_{cb}\vert}
\newcommand{\absvub}{\vert V_{ub}\vert}
\newcommand{\absvtd}{\vert V_{td}\vert}
\newcommand{\absvts}{\vert V_{ts}\vert}
\newcommand{\abseps}{\vert\epsilon\vert}

\newcommand{\fbb}{f^2_{B_d}\hat{B}_{B_d}}
\newcommand{\fbbs}{f^2_{B_s}\hat{B}_{B_s}}
\newcommand{\fbd}{f_{B_d}}
\newcommand{\fbs}{f_{B_s}}

\newread\epsffilein    
\newif\ifepsffileok    
\newif\ifepsfbbfound   
\newif\ifepsfverbose   
\newdimen\epsfxsize    
\newdimen\epsfysize    
\newdimen\epsftsize    
\newdimen\epsfrsize    
\newdimen\epsftmp      
\newdimen\pspoints     
\def\epsfbox#1{\global\def\epsfllx{72}\global\def\epsflly{72}%
   \global\def\epsfurx{540}\global\def\epsfury{720}%
   \def\lbracket{[}\def\testit{#1}\ifx\testit\lbracket
   \let\next=\epsfgetlitbb\else\let\next=\epsfnormal\fi\next{#1}}%
\def\epsfgetlitbb#1#2 #3 #4 #5]#6{\epsfgrab #2 #3 #4 #5 .\\%
   \epsfsetgraph{#6}}%
\def\epsfnormal#1{\epsfgetbb{#1}\epsfsetgraph{#1}}%
\def\epsfgetbb#1{%
\openin\epsffilein=#1
\ifeof\epsffilein\errmessage{I couldn't open #1, will ignore it}\else
   {\epsffileoktrue \chardef\other=12
    \def\do##1{\catcode`##1=\other}\dospecials \catcode`\ =10
    \loop
       \read\epsffilein to \epsffileline
       \ifeof\epsffilein\epsffileokfalse\else
          \expandafter\epsfaux\epsffileline:. \\%
       \fi
   \ifepsffileok\repeat
   \ifepsfbbfound\else
    \ifepsfverbose\message{No bounding box comment in #1; using defaults}\fi\fi
   }\closein\epsffilein\fi}%
\def\epsfclipstring{}
\def\epsfsetgraph#1{%
   \epsfrsize=\epsfury\pspoints
   \advance\epsfrsize by-\epsflly\pspoints
   \epsftsize=\epsfurx\pspoints
   \advance\epsftsize by-\epsfllx\pspoints
   \epsfxsize\epsfsize\epsftsize\epsfrsize
   \ifnum\epsfxsize=0 \ifnum\epsfysize=0
      \epsfxsize=\epsftsize \epsfysize=\epsfrsize
      \epsfrsize=0pt
     \else\epsftmp=\epsftsize \divide\epsftmp\epsfrsize
       \epsfxsize=\epsfysize \multiply\epsfxsize\epsftmp
       \multiply\epsftmp\epsfrsize \advance\epsftsize-\epsftmp
       \epsftmp=\epsfysize
       \loop \advance\epsftsize\epsftsize \divide\epsftmp 2
       \ifnum\epsftmp>0
          \ifnum\epsftsize<\epsfrsize\else
             \advance\epsftsize-\epsfrsize \advance\epsfxsize\epsftmp \fi
       \repeat
       \epsfrsize=0pt
     \fi
   \else \ifnum\epsfysize=0
     \epsftmp=\epsfrsize \divide\epsftmp\epsftsize
     \epsfysize=\epsfxsize \multiply\epsfysize\epsftmp
     \multiply\epsftmp\epsftsize \advance\epsfrsize-\epsftmp
     \epsftmp=\epsfxsize
     \loop \advance\epsfrsize\epsfrsize \divide\epsftmp 2
     \ifnum\epsftmp>0
        \ifnum\epsfrsize<\epsftsize\else
           \advance\epsfrsize-\epsftsize \advance\epsfysize\epsftmp \fi
     \repeat
     \epsfrsize=0pt
    \else
     \epsfrsize=\epsfysize
    \fi
   \fi
   \ifepsfverbose\message{#1: width=\the\epsfxsize, height=\the\epsfysize}\fi
   \epsftmp=10\epsfxsize \divide\epsftmp\pspoints
   \vbox to\epsfysize{\vfil\hbox to\epsfxsize{%
      \ifnum\epsfrsize=0\relax
        \includegraphics{#1}%
      \else
        \epsfrsize=10\epsfysize \divide\epsfrsize\pspoints
        \includegraphics{#1}%
      \fi
      \hfil}}%
\global\epsfxsize=0pt\global\epsfysize=0pt}%
 {\catcode`\%=12 \global\let\epsfpercent=
\long\def\epsfaux#1#2:#3\\{\ifx#1\epsfpercent
   \def\testit{#2}\ifx\testit\epsfbblit
      \epsfgrab #3 . . . \\%
      \epsffileokfalse
      \global\epsfbbfoundtrue
   \fi\else\ifx#1\par\else\epsffileokfalse\fi\fi}%
\def\epsfempty{}%
\def\epsfgrab #1 #2 #3 #4 #5\\{%
\global\def\epsfllx{#1}\ifx\epsfllx\epsfempty
      \epsfgrab #2 #3 #4 #5 .\\\else
   \global\def\epsflly{#2}%
   \global\def\epsfurx{#3}\global\def\epsfury{#4}\fi}%
\def\epsfsize#1#2{\epsfxsize}
\def\att{t \bar{t}}
\def\app{p \bar{p}}
\def\rts{\sqrt{s}}
\def\mt{m_t}
\def\mb{m_b}
\def\mc{m_c}

\newcommand{\delmd}{\Delta M_d}
\newcommand{\delms}{\Delta M_s}

\newcommand{\kkbar}{$K^0$-${\overline{K^0}}$}
\newcommand{\bdbdbar}{$B_d^0$-${\overline{B_d^0}}$}
\newcommand{\bsbsbar}{$B_s^0$-${\overline{B_s^0}}$}

\begin{document}
\begin{flushright}
CERN-TH.7398/94\\
UdeM-GPP-TH-94-06\\
\end{flushright}
\begin{center}
{\Large \bf
\centerline
{Implications of the Top Quark Mass Measurement}
\vspace*{0.3cm}
\centerline
{for the CKM Parameters, {\bf $\xs$} and CP Asymmetries}
\vspace*{0.3cm}
\centerline
{(Revised and Updated)}}
\vspace*{1.5cm}
 {\large A.~Ali}$\footnote{On leave of absence from DESY, Hamburg, FRG.}$
\vskip0.2cm
       Theory Division, CERN  \\
       CH-1211 Geneva 23, Switzerland \\
\vspace*{0.3cm}
\centerline{ and}
\vspace*{0.3cm}
\centerline{\large D.~London}
\smallskip
       Laboratoire de physique nucl\'eaire, Universit\'e de
Montr\'eal \\
             C.P. 6128, succ. centre-ville, Montr\'eal, QC, Canada
H3C 3J7\\
\vskip0.5cm
{\Large Abstract\\}
\parbox[t]{\textwidth}{
\indent
Motivated by the recent determination of the top quark mass by the CDF
collaboration, $\mt =174 \pm 10 ^{+13}_{-12}$ GeV, we review and update the
constraints on the parameters of the quark flavour mixing matrix $V_{CKM}$
in the standard model. In performing our fits, we use inputs from the
measurements of the following quantities: (i) $\abseps$, the CP-violating
parameter in $K$ decays, (ii) $\delmd$, the mass difference due to the
\bdbdbar\ mixing, (iii) the matrix elements $\absvcb$ and $\absvub$, and
(iv) $B$-hadron lifetimes. We find that the allowed region of the unitarity
triangle is very large, mostly due to theoretical uncertainties. (This
emphasizes the importance of measurements of CP-violating rate asymmetries
in the $B$ system.) Nevertheless, the present data do somewhat restrict the
allowed values of the coupling constant product
$f_{B_d}\sqrt{\hat{B}_{B_d}}$ and the renormalization-scale-invariant bag
constant $\hat{B}_K$. With the updated CKM matrix we present the
currently-allowed range of the ratio $\vert V_{td}/V_{ts} \vert$, as well
as the standard model predictions for the \bsbsbar\ mixing parameter $\xs$
and the quantities $\sin 2\alpha$, $\sin 2\beta$ and $\sin^2\gamma$, which
characterize the CP-asymmetries in $B$-decays. The ALEPH collaboration has
recently reported a significant improvement on the lower limit on the
$\bs$-$\bsb$ mass difference, $\Delta M_s/\Delta M_d > 11.3$ (95\% C.L.).
This has interesting consequences for the CKM parameters which are also
worked out.}
\vskip1cm
{\em Submitted to Zeitschrift f\"ur Physik C}
\end{center}
\noindent
CERN-TH.7398/94\\
August 1994
\thispagestyle{empty}
\newpage
\setcounter{page}{1}
\textheight 23.0 true cm


\section{Introduction}

The CDF collaboration at Fermilab has recently published evidence for top
quark production in $\app$ collisions at $\rts = 1.8$ TeV. The search is
based on the final states expected in the decays of the top quark in the
standard model (SM). Based on this analysis a top quark mass $\mt = 174 \pm
10 ^{+13}_{-12}$ GeV and a production cross section $\sigma (\app \to \att
+X)= 13.9^{+6.1}_{-4.8}~pb$ have been reported \cite{CDFmt}. The CDF value
for the top quark mass is in very comfortable agreement with the prediction
based on the SM fits of the electroweak data from LEP, SLC, CERN and
Fermilab colliders and neutrino beams, $\mt =177 \pm 11^{+18}_{-19}$ GeV
\cite{LEPew}. The top quark production cross section measured by CDF is
roughly a factor of 2 larger than the expected theoretical value in QCD
\cite{CDFmt} but is consistent with the upper limit presented by the D0
collaboration: $\sigma (\app \to \att +X) < 13~pb$ (95\% C.L.) for a top
quark mass of 180 GeV \cite{D0mt}. The neat overlap between the estimates
of $\mt$ based on the SM-electroweak analysis and its direct measurement,
together with the implied dominance of the decay mode $t \to W^+ b$, is a
resounding success of the standard model \cite{GSW,CKM}.

It is well appreciated that the top quark plays a crucial role in the
phenomenology of the electroweak interactions, flavour mixing, rare decay
rates and CP violation \cite{Burastop}. Therefore the new experimental
input for $\mt$, while still not very precise, should help in reducing the
present uncertainties on the parameters of the Cabibbo-Kobayashi-Maskawa
(CKM) quark mixing matrix \cite{CKM}. Conversely, the knowledge of $\mt$
can be used to restrict the range of the relevant hadronic matrix elements,
which in turn should help in firming up SM-based predictions for rare
decays and CP asymmetries in a number of $K$- and $B$-hadron decays. This
has been the theme of a number of papers which have appeared since the CDF
measurement of the top quark mass \cite{AL94,Pich94,deBoer94}. In addition,
a theoretical analysis along the same lines predated the CDF announcement
\cite{BLO94}. In the meantime, a number of parameters crucial for the CKM
phenomenology have evolved so that an updated analysis is in order.

The aim of this article is to revise and update the profile of the CKM
matrix elements reported earlier by us \cite{AL94}, in particular the CKM
unitarity triangle. In doing this update, we also include the improvements
reported in a number of measurements of the lifetime, mixing ratio, and the
CKM matrix elements $\absvcb$ and $\vert V_{ub}/V_{cb} \vert$ from $B$
decays, measured by the ARGUS, CLEO, CDF and LEP experiments. We note here
the changes that we have made in the present manuscript compared to our
earlier analysis reported in ref.~\cite{AL94}:
\begin{itemize}
\item
We have scaled the CDF top quark mass $\mt =174 \pm 16$ GeV to the running
top quark mass in the $\overline{MS}$ scheme so as to be able to use
consistently the next-to-leading order expressions for the mass differences
$\Delta M_d$ and $\Delta M_s$ and the CP violating parameter $\abseps$
which have been calculated in the $\overline{MS}$ scheme
\cite{Burastop,BLO94}. The correctly scaled value $\overline{\mt} (\mt)=
165 \pm 16 $ GeV \cite{mtmsbar} is about 9 GeV lower than what we had used
previously \cite{Pichthanks}. This renormalization reduces $\mt$ by only
$\sim 0.5\sigma$, and hence does not significantly change the results of
ref.~\cite{AL94}, though its incorporation into the CKM fits does change
the central values of the parameters slightly.
\item
The present knowledge of the CKM matrix element ratio $\vert V_{ub}/V_{cb}
\vert$ is poor. The only source of information for this ratio so far is the
end-point lepton energy spectrum in semileptonic $B$ decays which is
model dependent. No new measurement or analysis of this quantity has been
reported since we did our last fits. However, in consultation with our
experimental colleagues, we have increased the error on this ratio and now
use a value  $\vert V_{ub}/V_{cb} \vert = 0.08 \pm 0.03$, which better
reflects the underlying theoretical dispersion on this ratio.
\item
New measurements for the quantity
 ${\cal F}(1) \vert V_{cb} \vert$ in the
decays $B \to D^* \ell \nu_\ell$ using heavy quark effective theory (HQET)
methods have been made by the CLEO \cite{CLEOVcb94}, ARGUS
\cite{ARGUSVcb94} and ALEPH collaborations \cite{ALEPHVcb}. In the
meantime, estimates for the quantity ${\cal F}(1) \equiv \xi(1) \eta_A$
have undergone some revision in both the QCD perturbative part ($\eta_A)$
and power corrections to the Isgur-Wise function at the symmetry point
$(\xi(1)$) \cite{neubert94,suv94}. Taking into account the updated
experimental and theoretical input, we estimate $\vert V_{cb} \vert =0.039
\pm 0.006$. The central value for this matrix element has moved down by
$0.002$ compared to the value $\vert V_{cb} \vert =0.041 \pm 0.006$ used by
us previously.
\item
The precision on the $\bd$-$\bdb$ mass difference $\Delta M_d$ obtained
from time-dependent methods is now better than that on the corresponding
time-integrated quantity, $\xd =\Delta M_d/\Gamma_d$. The precision on
$\Delta M_d$ can be further improved by combining the time-dependent and
time-integrated information. The present world average, $\Delta M_d = 0.5
\pm 0.033$ (ps)$^{-1}$ \cite{Forty}, which includes the various systematic
uncertainties relevant for the extraction of this quantity, can be directly
analysed in terms of the (QCD-corrected) SM estimates of the same. This
reduces the error due to the lifetime, which one has to take into account
in the analysis of $\xd$.
\item
The lower limit on the mass difference ratio $\Delta M_s/\Delta M_d > 11.3$
at $95 \%$ C.L., reported by the ALEPH collaboration \cite{ALEPHxs},
provides an additional bound on the CKM parameters. Previously, the
experimental bound was much lower and hence not very interesting. In our
analysis, we present the constraints on the CKM parameters which follow
from the ALEPH lower limit on the mass difference ratio.
\end{itemize}

It should be pointed out that, despite these several changes, the results
presented in this paper differ very little from those of ref.~\cite{AL94}.
Furthermore, in spite of the benefit of new experimental data, the allowed
region for the CKM parameters is still quite large, as we will see. This is
because the main uncertainty is theoretical, and is due to our lack of
knowledge of hadronic matrix elements. This underlines the crucial
importance of measuring CP-violating rate asymmetries in the $B$ system --
such asymmetries are independent of hadronic uncertainties, and will thus
provide us with {\it clean} information about the CKM parameters.

In our analysis we consider two types of fits. In Fit 1, we assume
particular fixed values for the theoretical hadronic quantities. The
allowed ranges for the CKM parameters are derived from the (Gaussian)
errors on experimental measurements only. In Fit 2, we assign a central
value plus an error (treated as Gaussian) to the theoretical quantities. In
the resulting fits, we combine the experimental and theoretical errors in
quadrature. For both fits we calculate the allowed region in CKM parameter
space at 95\% C.L. We also present the corresponding allowed ranges for the
CP-violating phases that will be measured in $B$ decays, characterized by
$\sin 2\beta$, $\sin 2\alpha$ and $\sin^2\gamma$. These can be measured
directly through rate asymmetries in the decays $\bdbarp \to J/\psi K_S$,
$\bdbarp \to \pi^+ \pi^-$, and $\bsbarp\ \to D_s^\pm K^\mp$, respectively.
We also give the allowed domains for two of the angles, $(\sin 2\alpha,\sin
2\beta)$. We estimate the SM prediction for the \bsbsbar\ mixing parameter,
$\xs$, and show how the ALEPH limit of $\xs>9.0$ (95\% C.L.) constrains the
parameter space. Finally, we give the present $95\%$ C.L. upper and lower
bounds on the matrix element ratio $\vert V_{td}/V_{ts} \vert$.

This paper is organized as follows. In section 2, we present our update of
the CKM matrix, concentrating especially on the matrix element $\absvcb$
which, thanks to the progress in HQET and experiments, is now well under
control. The constraints that follow from $\vert V_{ub}/V_{cb} \vert$,
$\abseps$ and $\Delta M_d$ on the CKM parameters are also discussed here.
Section 3 contains the results of our fits. These results are summarized in
terms of the allowed domains of the unitarity triangle, which are displayed
in several Figures and Tables. In section 4, we discuss the impact of the
recent lower limit on the ratio $\Delta M_s/\Delta M_d$ reported by the
ALEPH collaboration on the CKM parameters and estimate the expected range
of the mixing ratio $\xs$ in the SM based on our fits. Here we also present
the allowed 95\% C.L. range for $\vert V_{td}/V_{ts} \vert$. In section 5
we discuss the predictions for the CP asymmetries in the neutral $B$ meson
sector and calculate the correlations for the CP violating asymmetries
proportional to $\sin 2\alpha$, $\sin 2 \beta$ and $\sin^2 \gamma$. Section
6 contains a summary and an outlook for improving the profile of the CKM
unitarity triangle.


\section{An Update of the CKM Matrix}

In updating the CKM matrix elements, we make use of the Wolfenstein
parametrization \cite{Wolfenstein}, which follows from the observation that
the elements of this matrix exhibit a hierarchy in terms of $\lambda$, the
Cabibbo angle. In this parametrization the CKM matrix can be written
approximately as
\beq
V_{CKM} \simeq \left(\matrix{
    1-{1\over 2}\lambda^2 & \lambda
                       & A\lambda^3 \left( \rho - i\eta \right) \cr
  -\lambda & 1-{1\over 2}\lambda^2 - i A^2 \lambda^4 \eta & A\lambda^2 \cr
   A\lambda^3\left(1 - \rho - i \eta\right) & -A\lambda^2 & 1 \cr}\right)~.
\label{CKM}
\eeq

In this section we shall discuss those quantities which constrain these CKM
parameters, pointing out the significant changes in the determination of
$\lambda$, $A$, $\rho$ and $\eta$. Recently, the importance of including
higher-order terms in the Wolfenstein parametrization given above has been
emphasized \cite{BLO94}. This amounts to redefining the parameters $\rho$
and $\eta$, with the improved Wolfenstein parameters being $\bar{\rho}=
\rho(1-\lambda^2/2)$ and $\bar{\eta}=\eta(1-\lambda^2/2)$. While such a
procedure may become important when the experimental precision on the CP
asymmetries in $B$ decays becomes comparable to $\lambda^2/2 \simeq 3\%$,
at present it is unnecessary, and we will continue using the standard
Wolfenstein parametrization given above. The error incurred by its use is
negligible compared to all the other uncertainties in the CKM fits.

We recall that $\vert V_{us}\vert$ has been extracted with good accuracy
from $K\to\pi e\nu$ and hyperon decays \cite{PDG} to be
\beq
\vert V_{us}\vert=\lambda=0.2205\pm 0.0018~.
\eeq
This agrees quite well with the determination of $V_{ud}\simeq 1-{1\over
2}\lambda^2$ from $\beta$-decay,
\beq
\vert V_{ud}\vert=0.9744\pm 0.0010~.
\eeq

The parameter $A$ is related to the CKM matrix element $V_{cb}$, which can
be obtained from semileptonic decays of $B$ mesons. We shall restrict
ourselves to the methods based on HQET to calculate the exclusive and
inclusive semileptonic decay rates. In the heavy quark limit it has been
observed that all hadronic form factors in the semileptonic decays $B \to
(D,D^*) \ell \nu_\ell$ can be expressed in terms of a single function, the
Isgur-Wise function \cite{Wisgur}. It has been shown that the HQET-based
method works best for $B\to D^*l\nu$ decays, since these are unaffected by
$1/m_Q$ corrections \cite{Luke,Boyd,Neubert}. This method has been used by
the ALEPH, ARGUS and CLEO collaborations to determine $\xi (1) \absvcb$ and
the slope of the Isgur-Wise function.

Using HQET, the differential decay rate in $B \to D^* \ell \nu_\ell$ is
\begin{eqnarray}
\frac{d\Gamma (B \to D^* \ell \bar{\nu})}{d\omega }
&=& \frac{G_F^2}{48 \pi^3} (m_B-m_{D^*})^2 m_{D^*}^3 \eta_{A}^2
  \sqrt{\omega^2-1} (\omega + 1)^2 \\ \nonumber
&~& ~~~~~~~~~~~~~~\times [ 1+ \frac{4 \omega}{\omega + 1}
 \frac{1-2\omega r + r^2}{(1-r)^2}] \absvcb ^2 \xi^2(\omega) ~,
\label{bdstara1}
\end{eqnarray}
where $r=m_{D^*}/m_B$, $\omega=v\cdot v'$ ($v$ and $v'$ are the
four-velocities of the $B$ and $D^*$ meson, respectively), and $\eta_{A}$
is the short-distance correction to the axial vector form factor. In the
leading logarithmic approximation, this was calculated by Shifman and
Voloshin some time ago -- the so-called hybrid anomalous dimension
\cite{hybrid}. In the absence of any power corrections, $\xi (\omega=1)=1$.
The size of the $O(1/\mb^2)$ and $O(1/\mc^2)$ corrections to the Isgur-Wise
function $\xi (\omega)$, and to some extent the next-to-leading order
corrections to $\eta_A$ have recently become a matter of some discussion
\cite{neubert94,suv94,neuberttasi}. We recall that the effects of such
power corrections were previously estimated as \cite{neuberttasi}
\begin{eqnarray}
\label{neubertxiold}
\xi (1) &=& 1+ \delta (1/m^2)= 0.98 \pm 0.04 ~, \nonumber \\
      \eta_{A} &=& 0.99 ~,
\end{eqnarray}
and the corresponding corrections to $\xi (1)$ were estimated by Mannel
\cite{MannelmQ} to be
\begin{equation}
\label{mannelxi}
 -0.05 < \xi (1) -1 < 0~.
\end{equation}
In a recent paper Shifman, Uraltsev and Vainshtain \cite{suv94} have argued
that the deviation of $\xi (1)$, as well as that of $\eta_A$, from unity is
larger than the estimate given in eq.~(\ref{neubertxiold}). Following
ref.~\cite{suv94}, this deviation can be expressed as
\beq
1-\xi^2(1) = \frac{1}{3}\frac{\mu_G^2}{m_c^2} + \frac{\mu_\pi^2-\mu_G^2}
{4} \big(\frac{1}{m_c^2} + \frac{1}{m_b^2} + \frac{2}{3 m_cm_b}\big)
+ \sum_{i=1,2,...} \xi^2_{B \to excit} ~,
\label{suveq1}
\eeq
where the contribution to the higher excited states is indicated by the
last term, and $\mu_G^2$ and $\mu_\pi^2$ parametrize the matrix elements of
the chromomagnetic and kinetic energy operators, respectively. These have
been estimated to be
\begin{eqnarray}
\mu_G^2 &=& \frac{3}{4} (M_{B^*}^2 - M_B^2) \simeq 0.35 ~\mbox{GeV}^2 ,
 \\ \nonumber
 \mu_\pi^2 &=&(0.54 \pm 0.12) ~\mbox{GeV}^2~,
\label{suveq2}
\end{eqnarray}
where the numbers for $\mu_\pi^2$ are based on QCD sum rules
\cite{ballbraun}. Using the central value for this quantity and ignoring
the contribution of the excited states, one gets (henceforth we use
${\cal F}(1)\equiv \eta_A \xi (1)$)
\beq
 {\cal F}(1)=0.92~.
\label{suveq3}
\eeq
The contribution of the higher states is positive definite. However, its
actual value can only be guessed at present. Shifman et al.\ estimate
\cite{suv94}
\beq
{\cal F}(1)=0.89 \pm 0.03~.
\label{suveq4}
\eeq
In the meantime, Neubert has revised his estimate of the same quantity (we
refer to ref.~\cite{neubert94} for details), obtaining
\beq
{\cal F}(1)=0.93 \pm 0.03~.
\label{neubertxinew}
\eeq

The values of ${\cal F}(1)$ given in eqs.~(\ref{suveq4}) and
(\ref{neubertxinew}) are significantly smaller than unity and one must
conclude that the $O(1/m_Q^2)$ corrections to $\xi(1)$ are important
numerically. However, the theoretical estimates  by Neubert and Shifman
et al. are now compatible with each other, within quoted errors. In the
analysis for $\absvcb$ presented here, we shall use the range
\beq
{\cal F}(1)=0.91 \pm 0.05~,
\label{alxi}
\eeq
which covers the $(\pm 1\sigma )$ range in the two estimates, giving an
intrinsic theoretical error $\Delta \vert V_{cb} \vert/\vert V_{cb} \vert
=0.06$ from corrections to ${\cal F}(1)$. To further reduce this error
a better estimate for the excited states is needed. This might be
forthcoming when the contribution of the inelastic channels in semileptonic
$B$ decays is measured more accurately.

As already mentioned in the introduction, the experimental measurements
have also evolved in time. The previously reported value for $\absvcb$ by
the ARGUS collaboration from the decays $B \to D^* + \ell \bar{\nu}$ using
the HQET formalism yielded a value $\absvcb =0.047 \pm 0.007$ with a
considerably higher value for the slope of the Isgur-Wise function, $\xi'
(1) \equiv -\rho^2$, in the range $1.9 < \rho^2 < 2.3$ \cite{argusvcb}. In
a recent analysis by ARGUS, significantly lower values for both $\absvcb$
and $\rho^2$ have been obtained, yielding \cite{ARGUSVcb94}
\begin{eqnarray}
{\cal F}(1) \absvcb \big(\frac{\tau(B_d^0)}{1.53 ~\mbox{ps}}\big)^{1/2}
 &=& 0.039\pm 0.004 \pm 0.003~,
\nonumber \\
 \rho^2 &=& 1.08 \pm 0.12~,
\label{newarguscb}
\end{eqnarray}
where the value of $\absvcb$ corresponds to a linear extrapolation of the
Isgur-Wise function $\xi(\omega) = 1-\rho^2(\omega -1)$, and the error
quoted includes that from the $B_d^0$ lifetime.

The numbers obtained by the CLEO collaboration using a similar method are
\cite{CLEOVcb94}
\begin{eqnarray}
{\cal F}(1) \absvcb \big(\frac{\tau(B_d^0)}{1.53 ~\mbox{ps}}\big)^{1/2}
 &=& 0.0351\pm 0.0019 \pm 0.0022~,
\nonumber \\
\rho^2 & = & 0.84 \pm 0.12 \pm 0.08~,
\label{newcleocb}
\end{eqnarray}
where the numbers correspond to a linear extrapolation of the Isgur-Wise
function. The error quoted includes also that from the $B$ lifetime,
$\tau(B_d^0)=1.53 \pm 0.09$ (ps). The slope of the Isgur-Wise function has
also been measured by the CLEO collaboration through an independent method
which yields in addition the ratios of the vector and axial vector form
factors in the decay $B \to D^* \ell/\nu_\ell$. The resulting slope
\cite{CLEOIW2},
\beq
\rho^2 = 1.01 \pm 0.15 \pm 0.09 ~,
\label{newcleoiw}
\eeq
is consistent with that given in eq.~(\ref{newcleocb}), though the central
value is larger. The third measurement along the same lines has been
undertaken by the ALEPH collaboration. Their analysis yields
\begin{eqnarray}
{\cal F}(1) \absvcb \big(\frac{\tau(B_d^0)}{1.53 ~\mbox{ps}}\big)^{1/2}
 &=& 0.0392\pm 0.0044 \pm 0.0035~,
\nonumber \\
\rho^2 & = & 0.46 \pm 0.30 \pm 0.15~.
\label{newalephcb}
\end{eqnarray}
All three measurements of the quantity ${\cal F}(1) \vert V_{cb} \vert$ are
compatible with one another. The slope parameter in eqs.~(\ref{newarguscb})
and (\ref{newcleoiw}) are very close to each other, with the ALEPH value
lower but consistent with the other two. All three measurements are in
agreement with the theoretical bounds, which suggest $\rho^2 \leq 1$
\cite{Bjorken,Voloshin}.

The ALEPH, ARGUS and CLEO values of ${\cal F}(1) \absvcb$ given above have
been averaged by taking into account various common and independent
systematic errors, yielding \cite{Casselpc}
\begin{equation}
{\cal F}(1) \absvcb
 = 0.036 \pm 0.003 ~.
\label{newf1vcb}
\end{equation}
Taking into account the theoretical estimate of the renormalized
Isgur-Wise function ${\cal F}(1)= 0.91 \pm 0.05$, the updated value for
$\absvcb$ can be expressed as
\beq
\absvcb = 0.0395 \pm 0.003 \pm 0.001 \pm 0.002 ~,
\label{ourvcb}
\eeq
where, following the advice in refs.~\cite{Casselpc,Patterson}, an overall
systematic error of $\pm 0.001$ has been added to take into account the
slope of the Isgur-Wise function; $\pm 0.002$ represents the theoretical
error due to ${\cal F}(1)$. Using the same data but taking instead the
estimate for ${\cal F}(1)$ in eq.~(\ref{neubertxinew}), Neubert has
obtained the following number for $\absvcb$ \cite{neubert94}:
\beq
\absvcb = 0.0399 \pm 0.0026(\mbox{exp}) \pm 0.0013 (\mbox{th}) ~,
\label{neubertvcb}
\eeq
getting $\absvcb =0.0399 \pm 0.0029$, adding all errors in quadrature. The
central values for $\absvcb$ in (\ref{ourvcb}) and (\ref{neubertvcb}) are
practically identical, but the associated errors are different. In our
opinion, it is perhaps prudent to add the errors linearly to have a
reliable determination of $\absvcb$. For the purposes of the fits which
follow, we shall use the following value of $\absvcb$:
\beq
\absvcb = 0.039 \pm 0.006 ~,
\label{alvcb}
\eeq
which yields for the CKM parameter $A$,
\beq
A = 0.80 \pm 0.12~.
\label{Avalue}
\eeq

The above value for $\absvcb$ is in broad agreement with the values
obtained using other theoretical techniques for the exclusive and inclusive
semileptonic decays. We mention here the lattice-QCD based calculation of
the Isgur-Wise function by the UKQCD collaboration which, when combined
with the CLEO and ARGUS data, yields \cite{UKQCD}
\beq
\absvcb = 0.037 \pm 0.001 \pm 0.002^{+0.008}_{-0.003}~,
\label{ukqcdvcb}
\eeq
where we have updated the published UKQCD number for the current value of
the $B$ lifetime. The updated number \cite{Sonipc} from  the lattice-based
calculation in ref.~\cite{BSSQCD} is
\beq
\absvcb = 0.044 \pm 0.007 \pm 0.005~.
\label{bssvcb}
\eeq
Likewise, a calculation using the QCD sum rules yields \cite{Narisonvcb}
\beq
\absvcb = 0.0382 \pm 0.0012  \pm 0.0015 \pm 0.0015~,
\label{narvcb}
\eeq
where the first error is experimental and the other two are
related to theoretical effects.
Inclusive semileptonic decay rates have also been calculated using HQET
methods \cite{suv94,LS94,ballnierste} which require the knowledge of the
quark mass difference $(\mb - \mc )$, the $B$ semileptonic branching ratio,
and the $B$ lifetime. There exists at present some dispersion in
the theoretical estimates of the quark masses and the mass difference,
which leads to significantly different determinations of the matrix element
$\absvcb$. The determination of $\absvcb$ using this method and data from
$\Upsilon (4S)$ and $Z$ decays is summarized in ref.~\cite{Patterson}:
\begin{eqnarray}
\Upsilon (4S): ~~~ \absvcb &=& 0.039 \pm 0.001 \pm 0.005 ~~~(\tau(B)=
1.63 \pm 0.07 ~\mbox{ps}),\nonumber \\
Z: ~~~~~~~~~ \absvcb &=& 0.042 \pm 0.002 \pm 0.005 ~~~(\tau(b)=
1.55 \pm 0.06~\mbox{ps}),
\label{vcbincl}
\end{eqnarray}
where the second error reflects theoretical dispersion. One notices a
remarkable consistency in $\absvcb$ from the $\Upsilon(4S)$ data using the
inclusive semileptonic rate and the corresponding number determined from
the exclusive decay $B \to D^* \ell \nu_\ell$ in the HQET approach, though
the present theoretical errors in the former are somewhat larger.

The other two CKM parameters $\rho$ and $\eta$ are constrained by the
measurements of $\vert V_{ub}/V_{cb}\vert$, $\abseps$ (the CP-violating
parameter in the kaon system), $\xd$ (\bdbdbar\ mixing) and (in principle)
$\epsilon^\prime/\epsilon$ ($\Delta S=1$ CP-violation in the kaon system).
We shall not discuss the constraints from $\epsilon^\prime/\epsilon$, due
to the various experimental and theoretical uncertainties surrounding it at
present, but take up the rest in turn and present fits in which the allowed
region of $\rho$ and $\eta$ is shown.

First of all, $\vert V_{ub}/V_{cb}\vert$ can be obtained by looking at the
endpoint of the inclusive lepton spectrum in semileptonic $B$ decays.
Unfortunately, there still exists quite a bit of model dependence in the
interpretation of data. The present average of this ratio, based on the
recent analysis of the ARGUS \cite{argusbu} and CLEO
\cite{cleo15bu,cleoIIbu} data, is \cite{Casselpc}
\beq
\left\vert \frac{V_{ub}}{V_{cb}} \right\vert = 0.08\pm 0.03~.
\label{vubvcbn}
\eeq
This gives
\beq
\sqrt{\rho^2 + \eta^2} = 0.36 \pm 0.14~.
\eeq
This is significantly less precise than the corresponding range used by us
\cite{AL94} and others \cite{Pich94,deBoer94,BLO94}, and has important
consequences for the CKM fits.

The experimental value of $\abseps$ is \cite{PDG}
\beq
\abseps = (2.26\pm 0.02)\times 10^{-3}~.
\eeq
Theoretically, $\abseps$ is essentially proportional to the imaginary part
of the box diagram for \kkbar\ mixing and is given by \cite{Burasetal}
\begin{eqnarray}
\abseps &=& \frac{G_F^2f_K^2M_KM_W^2}{6\sqrt{2}\pi^2\Delta M_K}
\hat{B}_K\left(A^2\lambda^6\eta\right)
\bigl(y_c\left\{\hat{\eta}_{ct}f_3(y_c,y_t)-\hat{\eta}_{cc}\right\}
 \nonumber \\
&~& ~~~~~~~~~~~~~~+ ~\hat{\eta}_{tt}y_tf_2(y_t)A^2\lambda^4(1-\rho)\bigr).
\label{eps}
\end{eqnarray}
Here, the $\hat{\eta}_i$ are QCD correction factors, of which
$\hat{\eta}_{cc}$ and $\hat{\eta}_{tt}$ have been calculated to
next-to-leading order, and, to the best of our knowledge,
$\hat{\eta}_{ct}$ has so far been calculated to leading order. The
renormalization-scale invariant coefficients have the value
$\hat{\eta}_{cc}\simeq 1.10$ \cite{HN94},
 $\hat{\eta}_{tt}\simeq 0.57$ \cite{etaB},
$\hat{\eta}_{ct}\simeq 0.36$ for $\Lambda_{QCD}=200$ MeV
\cite{Burastop,Flynn}. In eq.~(\ref{eps}), $y_i\equiv m_i^2/M_W^2$,
and the functions $f_2$ and $f_3$ are given by
\begin{eqnarray}
f_2(x) &=& \frac{1}{4} + \frac{9}{4}\frac{1}{(1-x)}
- \frac{3}{2}\frac{1}{(1-x)^2}
- \frac{3}{2} \frac{x^2\ln x}{(1-x)^3}~, \nonumber \\
f_3(x,y) &=& \ln \frac{y}{x} - \frac{3y}{4(1-y)}
\left( 1 + \frac{y}{1-y}\ln y\right).
\end{eqnarray}
(The above form for $f_3(x,y)$ is an approximation, obtained in the limit
$x\ll y$. For the exact expression, see ref.~\cite{InamiLim}.)

The final parameter in the expression for $\abseps$ is the
renormalization-scale independent parameter $\hat{B}_K$, which represents
our ignorance of the hadronic matrix element $\langle K^0 \vert
{({\overline{d}}\gamma^\mu (1-\gamma_5)s)}^2 \vert
{\overline{K^0}}\rangle$. The evaluation of this matrix element has been
the subject of much work. The earlier results are summarized in
ref.~\cite{AL92}. For a recent comparative study of the various
calculational techniques, in particular a critical review of the chiral
perturbation theory based estimates, see ref.~\cite{Pich94}, which
advocates a range $\hat{B}_K= 0.50 \pm 0.15$, termed as ``best estimates,"
and $\hat{B}_K=0.55 \pm 0.25$ as the so-called ``conservative choice,"
rendering this quantity uncertain to more than a
 factor of 2. We note here
that significant progress in lattice-QCD methods has been made in
determining $\hat{B}_K$ \cite{Shigemitsu}, based on a better understanding
of the perturbative corrections to lattice operators, finite lattice-size
effects, and first estimates of effects due to dynamical quarks. The
present lattice QCD estimate gives (Gupta et al.\ in
ref.~\cite{bklattice})
\beq
\hat{B}_K=0.82 \pm 0.027 \pm 0.023~.
\eeq

In our first set of fits, we consider specific values in the range 0.4 to
1.0 for $\hat{B}_K$. As we shall see, for $\hat{B}_K = 0.4$ a poor fit to
the data is obtained, so that such small values are somewhat disfavoured.
In Fit 2, we assign a central value plus an error to $\hat{B}_K$. In order
to reflect the estimates of this quantity in lattice QCD \cite{bklattice},
we take
\beq
\hat{B}_K = 0.8 \pm 0.2 ~.
\label{BKrange1}
\eeq
In order to see the sensitivity of the allowed range of the CKM parameters
to a lower value of $\hat{B}_K$, we also present fits for the range
\beq
\hat{B}_K = 0.6 \pm 0.2 ~,
\label{BKrange2}
\eeq
which overlaps with the values suggested by chiral perturbation theory. As
we will see, there is not an enormous difference in the results for the two
ranges.

We now turn to \bdbdbar\ mixing. The present world average of $\xd\equiv
\Delta M_d/\Gamma_d$, which is a measure of this mixing, is \cite{Forty}
\beq
\xd = 0.76 \pm 0.06~,
\label{xdvalue}
\eeq
which is based on time-integrated measurements which directly measure
$\xd$, and on time-dependent measurements which measure the mass difference
$\Delta M_d$ directly. This is then converted to $\xd$ using the $B_d^0$
lifetime. From a theoretical point of view it is better to use the mass
difference $\Delta M_d$, as it liberates one from the errors on the
lifetime measurement. In fact, the present precision on $\Delta M_d$,
pioneered by time-dependent techniques at LEP, is quite competitive with
the precision on $\xd$. The LEP-average $\Delta M_d= 0.513 \pm 0.036$
(ps)$^{-1}$ has been combined with that derived from time-integrated
measurements yielding the present world average \cite{Forty}
\beq
\Delta M_d = 0.500 \pm 0.033 ~\mbox{(ps)}^{-1} ~.
\label{deltamd}
\eeq
We shall use this number instead of $\xd$, which has been the usual
practice to date \cite{Burastop}-\cite{BLO94}.

\begin{table}
\hfil
\vbox{\offinterlineskip
\halign{&\vrule#&
   \strut\quad#\hfil\quad\cr
\noalign{\hrule}
height2pt&\omit&&\omit&\cr
& Parameter && Value & \cr
height2pt&\omit&&\omit&\cr
\noalign{\hrule}
height2pt&\omit&&\omit&\cr
&  $\lambda$ && $0.2205$ & \cr
&  $\vert V_{cb} \vert $ && $0.039 \pm 0.006$ & \cr
&  $\vert V_{ub} / V_{cb} \vert$  && $0.08 \pm 0.03$ & \cr
&  $\abseps$  && $(2.26 \pm 0.02) \times 10^{-3}$ & \cr
&  $\Delta M_d$ && $(0.50 \pm 0.033)$ (ps)$^{-1}$  & \cr
&  $\overline{\mt}(\mt(pole))$ && $(165 \pm 16)$ GeV & \cr
&  $\hat{\eta}_B$  && $0.55$ & \cr
&  $\hat{\eta}_{cc} $ && $1.10$ & \cr
&  $\hat{\eta}_{ct} $ && $0.36$ & \cr
&  $\hat{\eta}_{tt} $ && $0.57$ & \cr
&  $\hat{B}_K$ && $0.8 \pm 0.2$ & \cr
&  $\hat{B}_B$ && $1.0 \pm 0.2$ & \cr
&  $\fbd$ && $180 \pm 50$ MeV & \cr
height2pt&\omit&&\omit&\cr
\noalign{\hrule}}}
\caption{Parameters used in the CKM fits. Values of the hadronic quantities
$\fbd$, $\hat{B}_{B_d}$ and $\hat{B}_K$ shown are motivated by the lattice
QCD results. In Fit 1, specific values of these hadronic quantities are
chosen, while in Fit 2, they are allowed to vary over the given ranges. (In
Fit 2, for comparison we also consider the range $\hat{B}_K = 0.6 \pm 0.2$,
which is motivated by chiral perturbation theory and QCD sum rules.)}
\label{tabfit}
\end{table}

The mass difference $\Delta M_d$ is calculated from the \bdbdbar\ box
diagram. Unlike the kaon system, where the contributions of both the $c$-
and the $t$-quarks in the loop were important, this diagram is dominated by
$t$-quark exchange:
\beq
\label{bdmixing}
\Delta M_d = \frac{G_F^2}{6\pi^2}M_W^2M_B\left(\fbb\right)\hat{\eta}_B y_t
f_2(y_t) \vert V_{td}^*V_{tb}\vert^2~, \label{xd}
\eeq
where, using eq.~\ref{CKM}, $\vert V_{td}^*V_{tb}\vert^2=
A^2\lambda^{6}\left[\left(1-\rho\right)^2+\eta^2\right]$. Here,
$\hat{\eta}_B$ is the QCD correction. In ref.~\cite{etaB}, this correction
is analyzed including the effects of a heavy $t$-quark. It is found that
$\hat{\eta}_B$ depends sensitively on the definition of the $t$-quark mass,
and that, strictly speaking, only the product $\hat{\eta}_B(y_t)f_2(y_t)$
is free of this dependence. In the fits presented here we use the value
$\hat{\eta}_B=0.55$, calculated in the $\overline{MS}$ scheme, following
ref.~\cite{etaB}. Consistency requires that the top quark mass be rescaled
from its pole (mass) value of $\mt =174 \pm 16$ GeV to the value
$\overline{\mt}(\mt(pole))$ in the $\overline{MS}$ scheme, which is
typically about 9 GeV smaller \cite{mtmsbar,Pichthanks}.

For the $B$ system, the hadronic uncertainty is given by $\fbb$, analogous
to $\hat{B}_K$ in the kaon system, except that in this case, also $\fbd$ is
not measured. The lattice-QCD results for these hadronic quantities have
been recently summarized as follows \cite{Shigemitsu}:
\begin{eqnarray}
f_{B_d} &=& 180 \pm 40 ~\mbox{MeV} ~, \nonumber \\
\hat{B}(B_d) & = & 1.2 \pm 0.2 ~,
\end{eqnarray}
while the corresponding numbers from the QCD sum rules are \cite{Narison}
\begin{eqnarray}
f_{B_d} &=& (1.60 \pm 0.26) f_\pi ~, \nonumber \\
\hat{B}(B_d) & = & 1.0 \pm 0.15 ~.
\end{eqnarray}
In our fits, we will take ranges for $\fbb$ and $\hat{B}_{B_d}$ which are
compatible with both of the above sets of numbers
\cite{Shigemitsu,Narison,Sommer94}:
\begin{eqnarray}
\fbd &=& 180 \pm 50 ~\mbox{MeV}~, \nonumber \\
\hat{B}_{B_d} &=& 1.0 \pm 0.2 ~.
\label{FBrange}
\end{eqnarray}
In Table \ref{tabfit}, we summarize all input quantities to our fits.


\section{The Unitarity Triangle}

\begin{figure}
\centerline{\psfig{figure=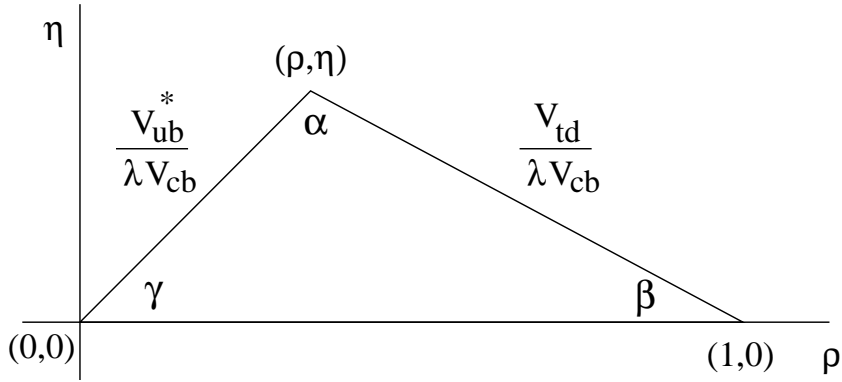,height=5.0cm,angle=90}}
\caption{The unitarity triangle. The angles $\alpha$, $\beta$ and $\gamma$
can be measured via CP violation in the $B$ system.}
\label{triangle}
\end{figure}

The allowed region in $\rho$-$\eta$ space can be displayed quite elegantly
using the so-called unitarity triangle. The unitarity of  the CKM matrix
leads to the following relation:
\beq
V_{ud} V_{ub}^* + V_{cd} V_{cb}^* + V_{td} V_{tb}^* = 0~.
\eeq
Using the form of the CKM matrix in eq.~\ref{CKM}, this can be recast as
\beq
\frac{V_{ub}^*}{\lambda V_{cb}} + \frac{V_{td}}{\lambda V_{cb}} = 1~,
\eeq
which is a triangle relation in the complex plane (i.e.\ $\rho$-$\eta$
space), illustrated in Fig.~\ref{triangle}. Thus, allowed values of $\rho$
and $\eta$ translate into allowed shapes of the unitarity triangle.

In order to find the allowed unitarity triangles, the computer program
MINUIT is used to fit the CKM parameters $A$, $\rho$ and $\eta$ to the
experimental values of $\absvcb$, $\vert V_{ub}/V_{cb}\vert$, $\abseps$ and
$\xd$. Since $\lambda$ is very well measured, we have fixed it to its
central value given above. As discussed in the introduction, we present
here two types of fits:
\begin{itemize}
\item
Fit 1: the ``experimental fit.'' Here, only the experimentally measured
numbers are used as inputs to the fit with Gaussian errors; the coupling
constants $f_{B_d} \sqrt{\hat{B}_{B_d}}$ and $\hat{B}_K$ are given fixed
values.
\item
Fit 2: the ``combined fit.'' Here, both the experimental and theoretical
numbers are used as inputs assuming Gaussian errors for the theoretical
quantities.
\end{itemize}

We first discuss the ``experimental fit" (Fit 1). The goal here is to
restrict the allowed range of the parameters ($\rho,\eta)$ for given values
of the coupling constants $f_{B_d} \sqrt{\hat{B}_{B_d}}$ and $\hat{B}_K$.
For each value of $\hat{B}_K$ and $f_{B_d}\sqrt{\hat{B}_{B_d}}$, the CKM
parameters $A$, $\rho$ and $\eta$ are fit to the experimental numbers given
in Table \ref{tabfit} and the $\chi^2$ is calculated.

\begin{figure}
\vskip -2.4truein
\centerline{\epsfxsize 2.6 truein \epsfbox {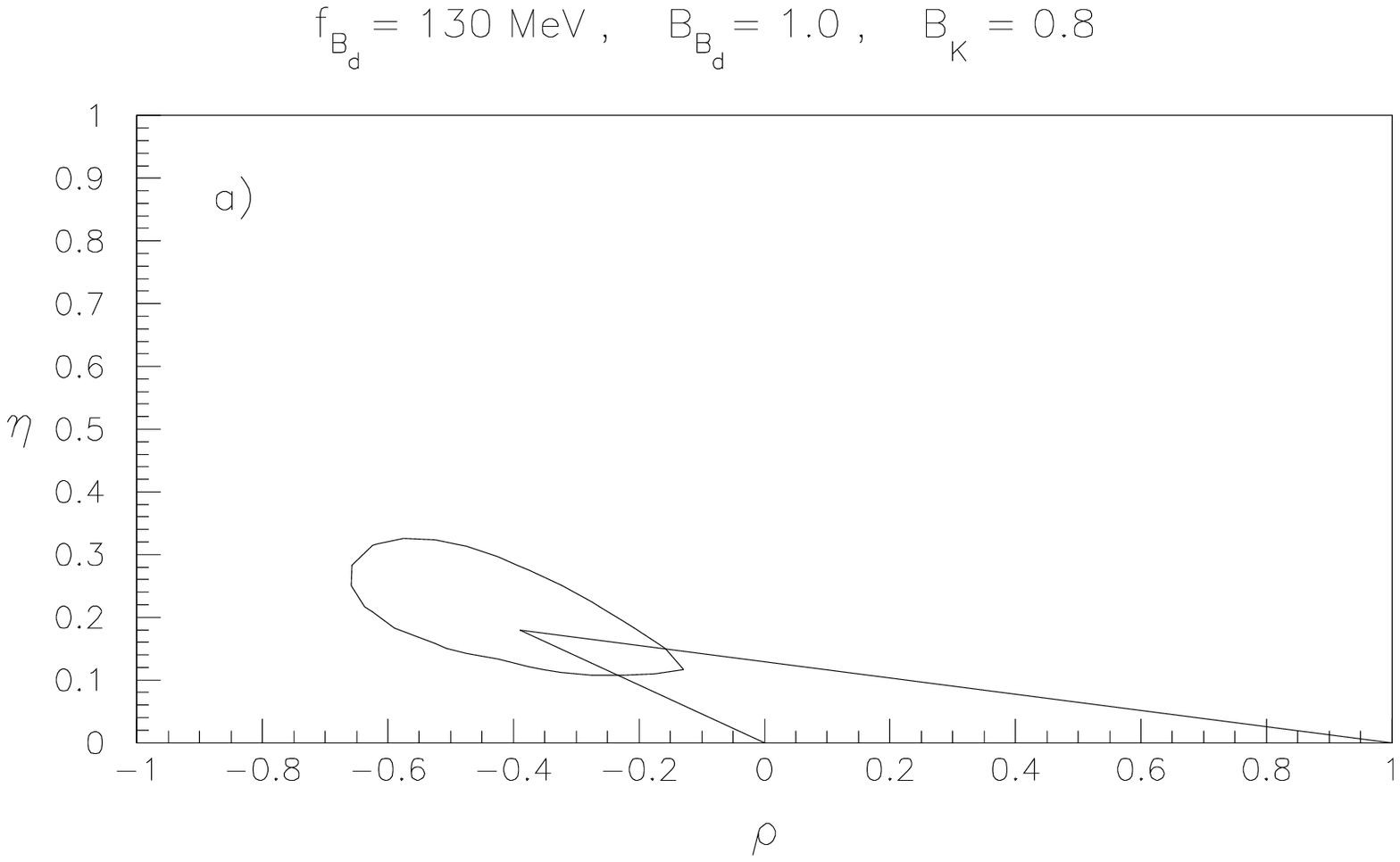}}
\vskip -1.8truein
\centerline{\epsfxsize 2.6 truein \epsfbox {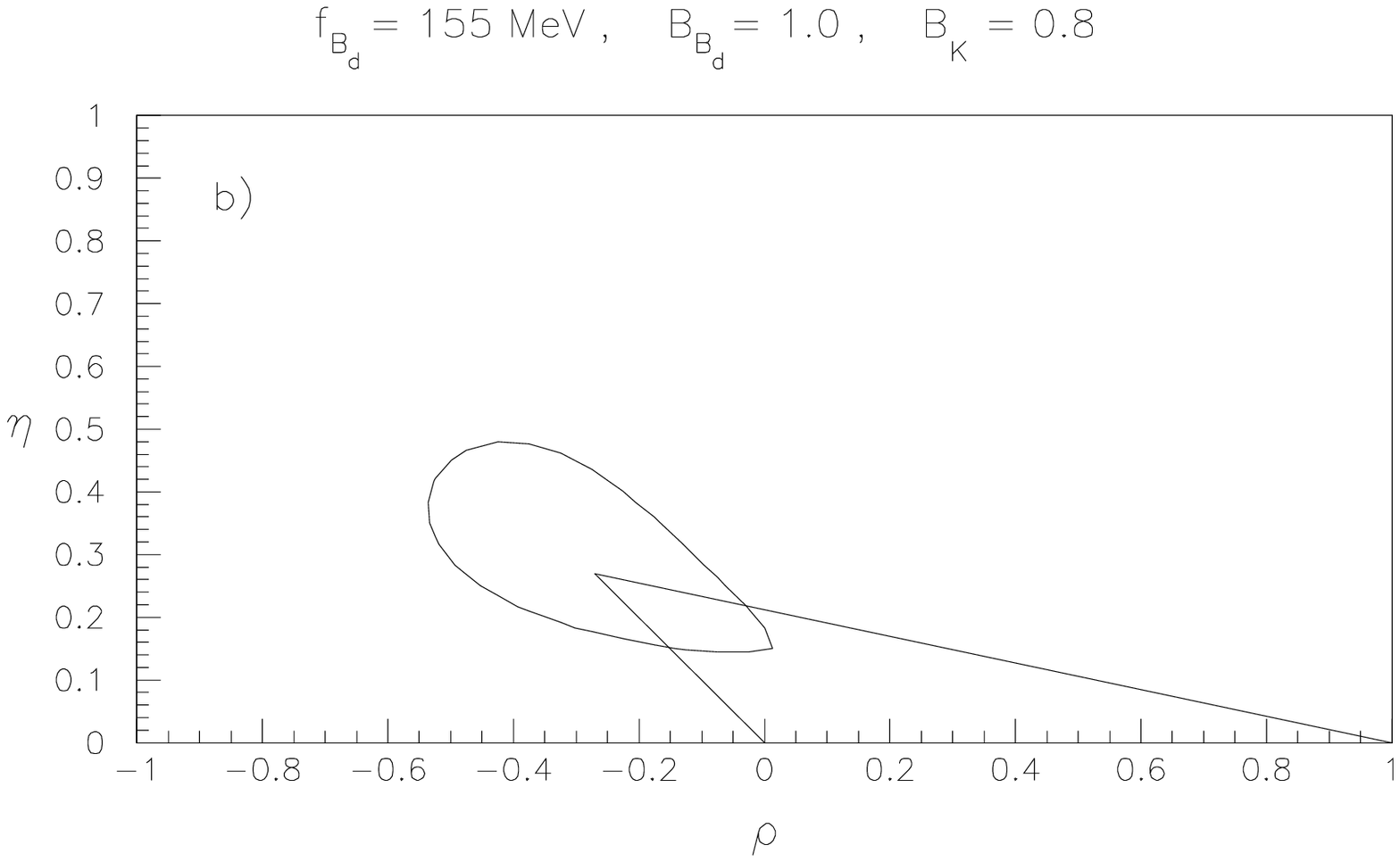}}
\vskip -1.8truein
\centerline{\epsfxsize 2.6 truein \epsfbox {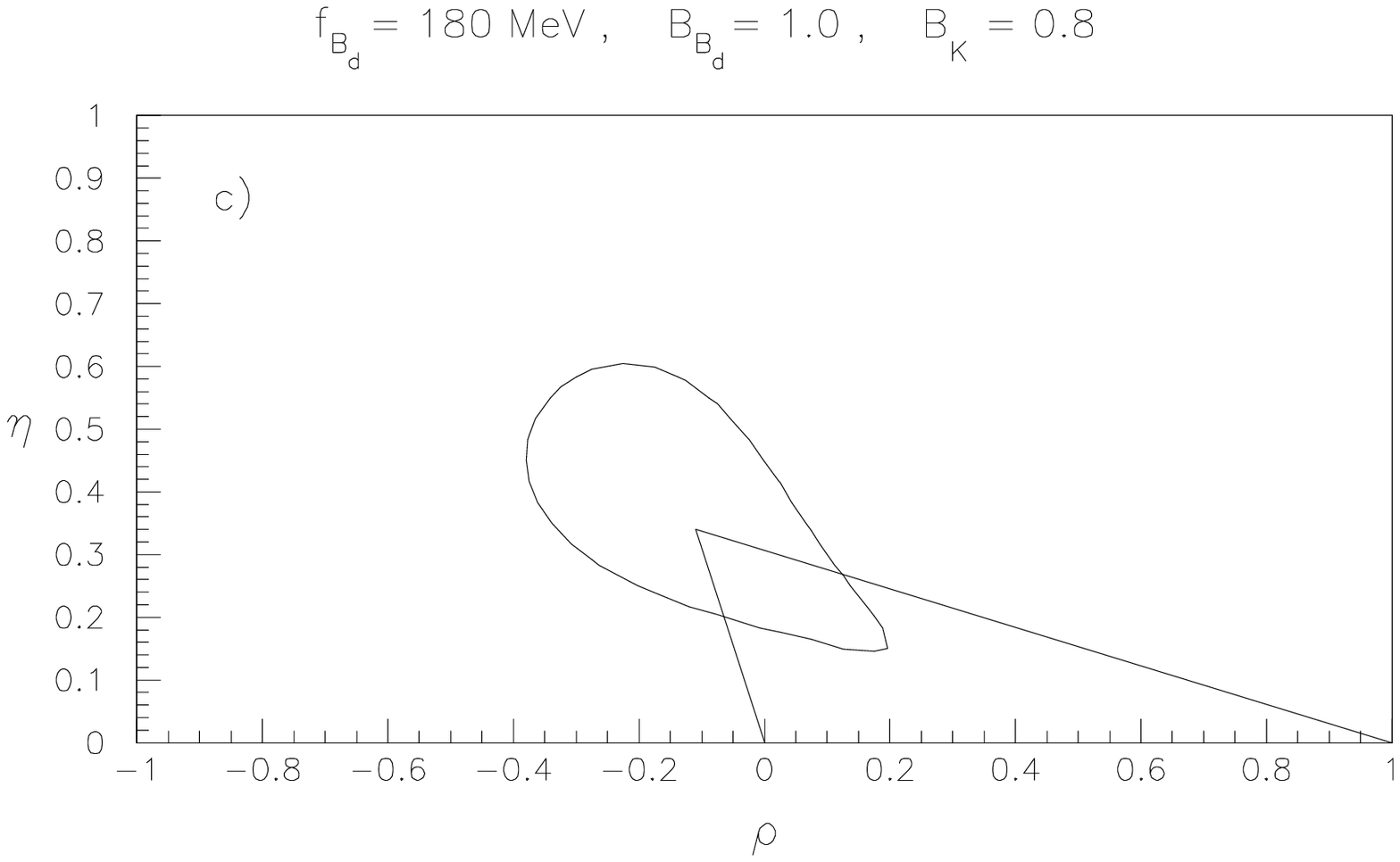}}
\vskip -1.8truein
\centerline{\epsfxsize 2.6 truein \epsfbox {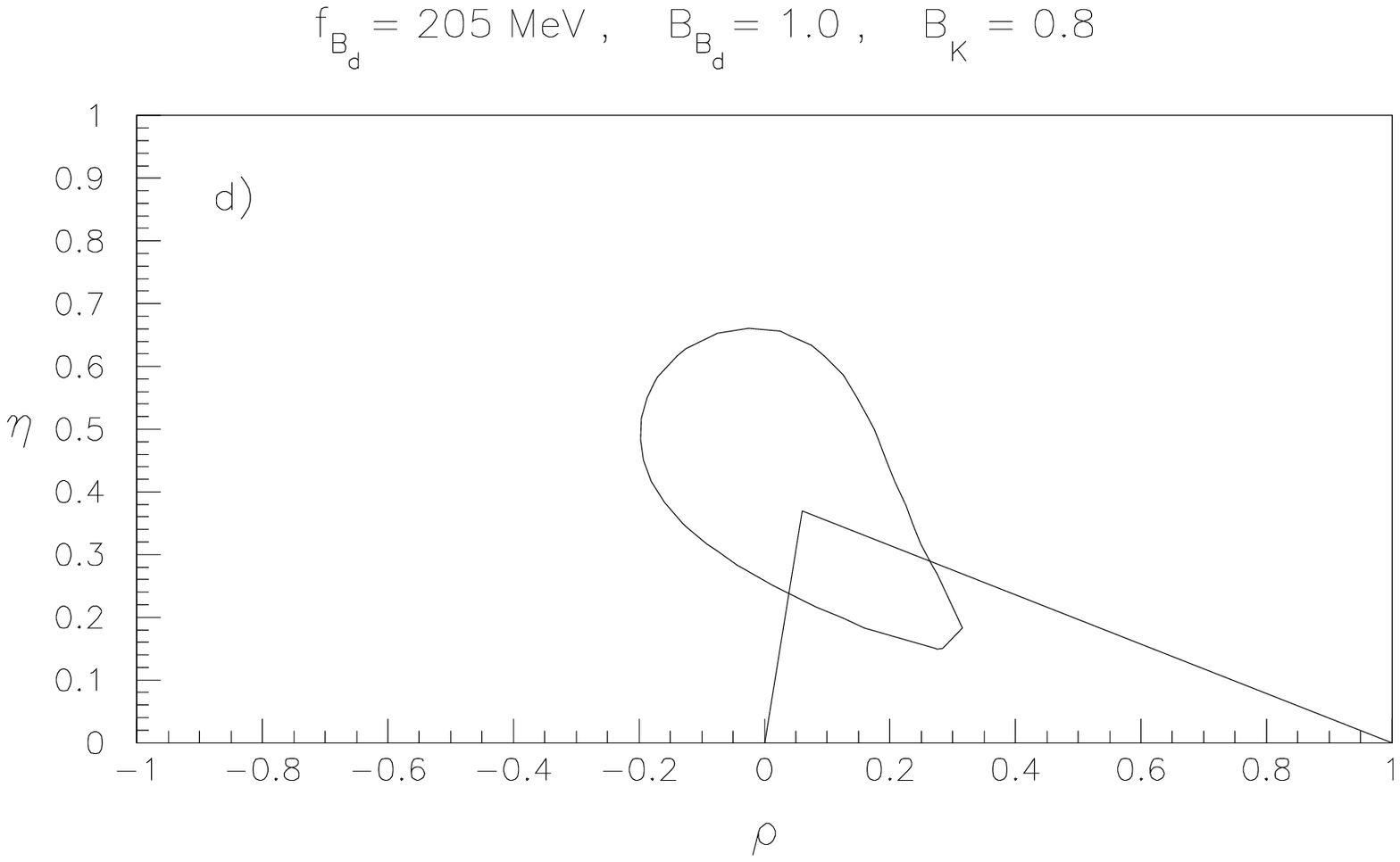}}
\vskip -1.8truein
\centerline{\epsfxsize 2.6 truein \epsfbox {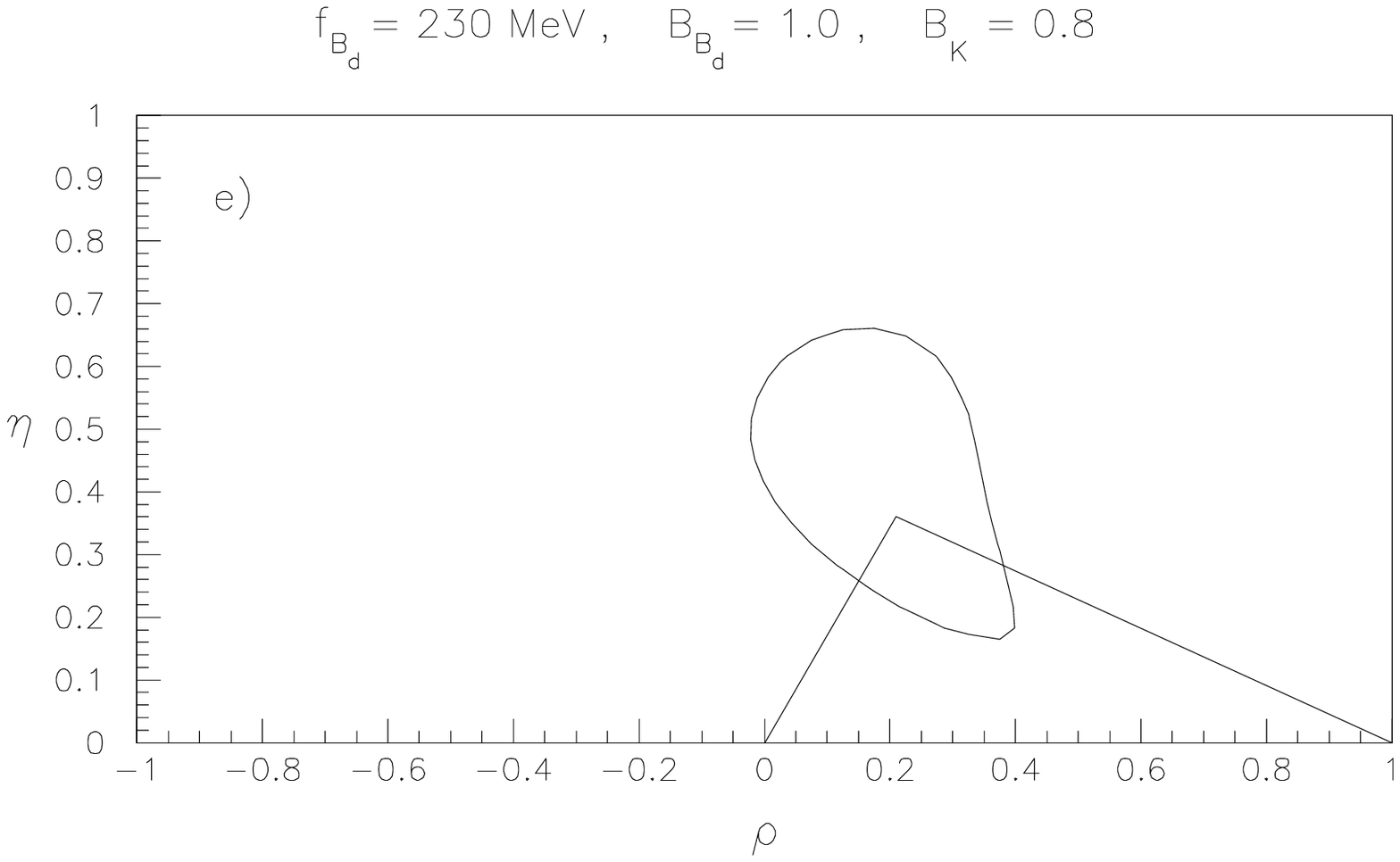}}
\vskip -1.0truein
\caption{Allowed region in $\rho$-$\eta$ space, from a fit to the
experimental values given in Table \protect{\ref{tabfit}}. We have fixed
$\hat{B}_K=0.8$ and vary the coupling constant product
$\fbd\protect\sqrt{\hat{B}_{B_d}}$ as indicated on the figures. The solid
line represents the region with $\chi^2=\chi_{min}^2+6$ corresponding to
the 95\% C.L.\ region. The triangles show the best fit.}
\label{rhoeta1}
\end{figure}

\begin{figure}
\vskip -2.2truein
\centerline{\epsfxsize 3.5 truein \epsfbox {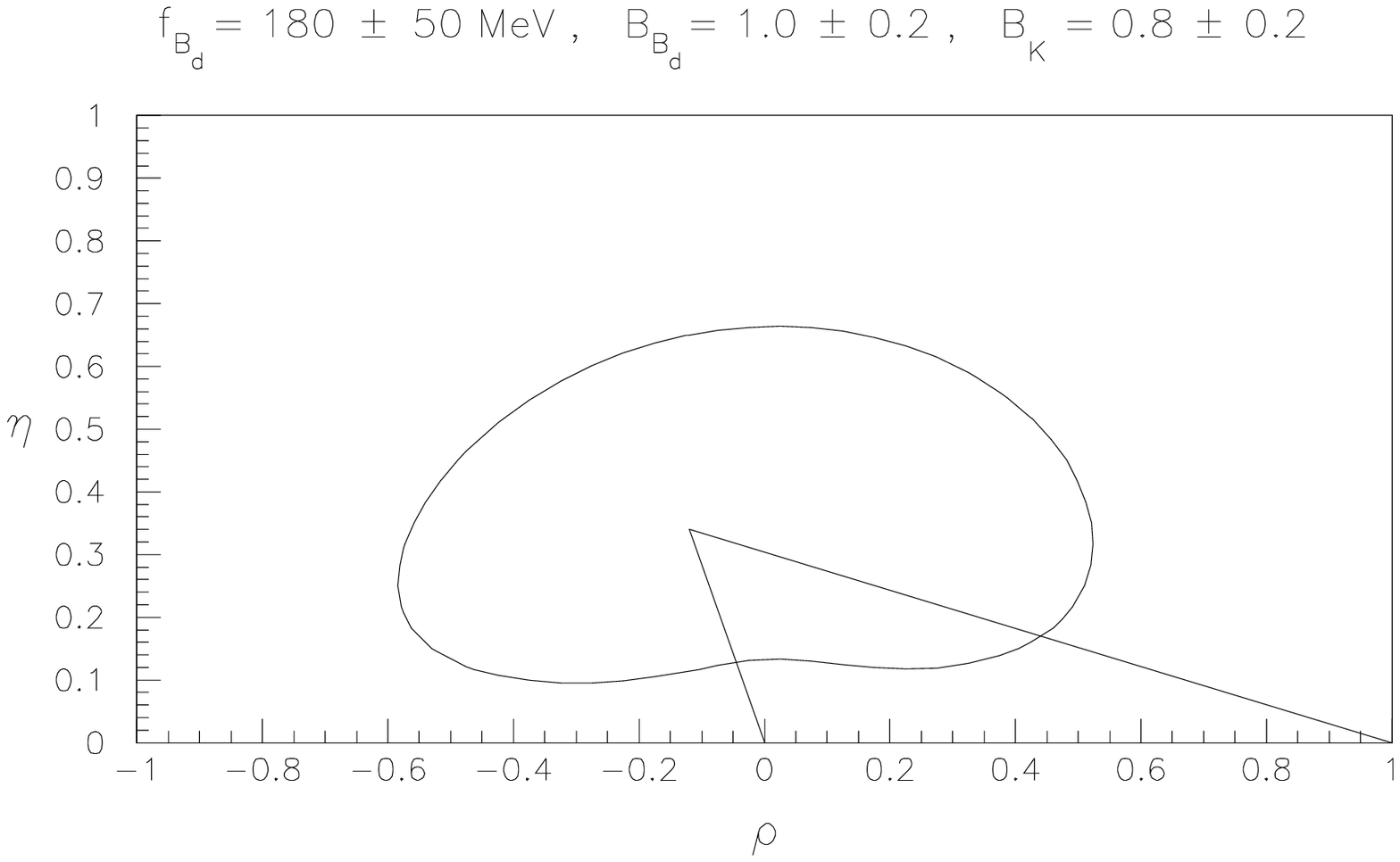}}
\vskip -1.4truein
\caption{Allowed region in $\rho$-$\eta$ space, from a simultaneous fit to
both the experimental and theoretical quantities given in Table
\protect{\ref{tabfit}}. The theoretical errors are treated as Gaussian for
this fit. The solid line represents the region with $\chi^2=\chi_{min}^2+6$
corresponding to the 95\% C.L.\ region. The triangle shows the best fit.}
\label{rhoeta2}
\end{figure}

First, we fix $\hat{B}_K = 0.8$, and vary $f_{B_d}\sqrt{\hat{B}_{B_d}}$ in
the range 130 MeV to 230 MeV. The fits are presented as an allowed region
in $\rho$-$\eta$ space at 95\% C.L. ($\chi^2 = \chi^2_{min} + 6.0$). The
results are shown in Fig.~\ref{rhoeta1}. As we pass from
Fig.~\ref{rhoeta1}(a) to Fig.~\ref{rhoeta1}(e), the unitarity triangles
represented by these graphs become more and more obtuse. Even more striking
than this, however, is the fact that the range of possibilities for these
triangles is enormous. There are two things to be learned from this. First,
our knowledge of the unitarity triangle is at present rather poor. This
will be seen even more clearly when we present the results of Fit 2.
Second, unless our knowledge of hadronic matrix elements improves
considerably, measurements of $\abseps$ and $x_d$, no matter how precise,
will not help much in further constraining the unitarity triangle. This is
why measurements of CP-violating rate asymmetries in the $B$ system are so
important \cite{BCPasym,AKL94}. Being largely independent of theoretical
uncertainties, they will allow us to accurately pin down the unitarity
triangle. With this knowledge, we could deduce the correct values of
$\hat{B}_K$ and $f_{B_d}\sqrt{\hat{B}_{B_d}}$, and thus rule out or confirm
different theoretical approaches to calculating these hadronic quantities.

Despite the large allowed region in the $\rho$-$\eta$ plane, certain values
of $\hat{B}_K$ and $f_{B_d}\sqrt{\hat{B}_{B_d}}$ are disfavoured since they
do not provide a good fit to the data. For example, fixing $\hat{B}_K=1.0$,
we can use the fitting program to provide the minimum $\chi^2$ for various
values of $f_{B_d}\sqrt{\hat{B}_{B_d}}$. The results are shown in Table
\ref{tabbk1}, along with the best fit values of $(\rho,\eta)$. Since we
have two variables ($\rho$ and $\eta$), we use $\chi^2_{min}<2.0$ as our
``good fit" criterion, and we see that $f_{B_d} \sqrt{\hat{B}_{B_d}} < 120$
MeV and $f_{B_d} \sqrt{\hat{B}_{B_d}} > 290$ MeV give poor fits to the
existing data. Note also that the $\chi^2$ distribution has two minima, at
around $f_{B_d} \sqrt{\hat{B}_{B_d}} = 160$ and 230 MeV. We do not consider
this terribly significant, since the surrounding values of $f_{B_d}
\sqrt{\hat{B}_{B_d}}$ also yield good fits to the data. The very small
values of $\chi^2_{min}$ depend sensitively on the central values of the
various experimental quantities -- if these values move around a little
bit, the values of $f_{B_d} \sqrt{\hat{B}_{B_d}}$ which give the minimum
$\chi^2$ values will move around as well. In Tables \ref{tabbk8},
\ref{tabbk6} and \ref{tabbk4}, we present similar analyses, but for
$\hat{B}_K=0.8$, $0.6$ and $0.4$, respectively. From these tables we see
that the lower limit on $f_{B_d} \sqrt{\hat{B}_{B_d}}$ remains fairly
constant, at around 120 MeV, but the upper limit depends quite strongly on
the value of $\hat{B}_K$ chosen. Specifically, for $\hat{B}_K =0.8$, $0.6$
and $0.4$, the maximum allowed value of $f_{B_d} \sqrt{\hat{B}_{B_d}}$ is
about 270, 230 and 180 MeV, respectively. Note that, for the lower value
$\hat{B}_K=0.4$, which is not favoured by lattice calculations or QCD sum
rules, the allowed range of $f_{B_d} \sqrt{\hat{B}_{B_d}}$ is quite
restricted, with generally higher values of $\chi^2$ than for the cases of
$\hat{B}_K$ in the range 0.6-1.0. This suggests that the data disfavour
(though do not completely exclude) $\hat{B}_K \leq 0.4$ solutions. Summing
up, present data exclude all values of $f_{B_d}\sqrt{f_{B_d}}$ which lie
below 110 MeV and above 290 MeV for the entire $\hat{B}_K$ range.

\begin{figure}
\centerline{\epsfxsize 3.5 truein \epsfbox {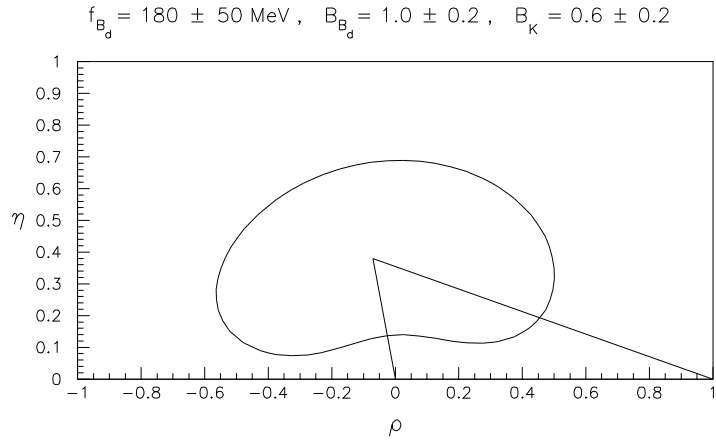}}
\vskip -1.4truein
\caption{Allowed region in $\rho$-$\eta$ space, from a simultaneous fit to
both the experimental and theoretical quantities given in Table
\protect{\ref{tabfit}}, except that we take $\hat{B}_K = 0.6 \pm 0.2$. The
theoretical errors are treated as Gaussian for this fit. The solid line
represents the region with $\chi^2=\chi_{min}^2+6$ corresponding to the
95\% C.L.\ region. The triangle shows the best fit.}
\label{rhoeta3}
\end{figure}

\begin{table}
\hfil
\vbox{\offinterlineskip
\halign{&\vrule#&
   \strut\quad#\hfil\quad\cr
\noalign{\hrule}
height2pt&\omit&&\omit&&\omit&\cr
& $\fbd\sqrt{\hat{B}_{B_d}}$ (MeV)
 && $(\rho,\eta)$ && $\chi^2_{min}$ & \cr
height2pt&\omit&&\omit&&\omit&\cr
\noalign{\hrule}
height2pt&\omit&&\omit&&\omit&\cr
& $110$ && $(-0.48,~0.10)$ && $3.24$ & \cr
& $120$ && $(-0.44,~0.12)$ && $1.77$ & \cr
& $130$ && $(-0.40,~0.15)$ && $0.85$ & \cr
& $140$ && $(-0.36,~0.18)$ && $0.33$ & \cr
& $150$ && $(-0.32,~0.21)$ && $7.6\times 10^{-2}$ & \cr
& $160$ && $(-0.28,~0.24)$ && $1.1\times 10^{-3}$ & \cr
& $170$ && $(-0.23,~0.27)$ && $2.4\times 10^{-2}$ & \cr
& $180$ && $(-0.17,~0.29)$ && $8.0\times 10^{-2}$ & \cr
& $190$ && $(-0.11,~0.32)$ && $0.12$ & \cr
& $200$ && $(-0.04,~0.33)$ && $0.13$ & \cr
& $210$ && $(0.03,~0.33)$ && $8.5\times 10^{-2}$ & \cr
& $220$ && $(0.09,~0.33)$ && $2.8\times 10^{-2}$ & \cr
& $230$ && $(0.15,~0.33)$ && $4.5\times 10^{-5}$ & \cr
& $240$ && $(0.21,~0.33)$ && $4.4\times 10^{-2}$ & \cr
& $250$ && $(0.25,~0.33)$ && $0.18$ & \cr
& $260$ && $(0.29,~0.33)$ && $0.43$ & \cr
& $270$ && $(0.33,~0.33)$ && $0.77$ & \cr
& $280$ && $(0.37,~0.33)$ && $1.21$ & \cr
& $290$ && $(0.40,~0.33)$ && $1.73$ & \cr
& $300$ && $(0.43,~0.32)$ && $2.34$ & \cr
height2pt&\omit&&\omit&&\omit&\cr
\noalign{\hrule}}}
\caption{The ``best values'' of the CKM parameters $(\rho,\eta)$ as a
function of the coupling constant $\fbd\protect\sqrt{\hat{B}_{B_d}}$,
obtained by a minimum $\chi^2$ fit to the experimental data, including the
renormalized value of $m_t=165 \pm 16$ GeV. We fix $\hat{B}_K=1.0$. The
resulting minimum $\chi^2$ values from the MINUIT fits are also given.}
\label{tabbk1}
\end{table}

\begin{table}
\hfil
\vbox{\offinterlineskip
\halign{&\vrule#&
   \strut\quad#\hfil\quad\cr
\noalign{\hrule}
height2pt&\omit&&\omit&&\omit&\cr
& $\fbd\sqrt{\hat{B}_{B_d}}$
 (MeV) && $(\rho,\eta)$ && $\chi^2_{min}$ & \cr
height2pt&\omit&&\omit&&\omit&\cr
\noalign{\hrule}
height2pt&\omit&&\omit&&\omit&\cr
& $110$ && $(-0.47,~0.12)$ && $3.29$ & \cr
& $120$ && $(-0.43,~0.15)$ && $1.83$ & \cr
& $130$ && $(-0.39,~0.18)$ && $0.92$ & \cr
& $140$ && $(-0.34,~0.22)$ && $0.40$ & \cr
& $150$ && $(-0.30,~0.25)$ && $0.13$ & \cr
& $160$ && $(-0.24,~0.29)$ && $2.7\times 10^{-2}$ & \cr
& $170$ && $(-0.18,~0.32)$ && $8.6\times 10^{-4}$ & \cr
& $180$ && $(-0.11,~0.34)$ && $1.3\times 10^{-3}$ & \cr
& $190$ && $(-0.04,~0.36)$ && $9.2\times 10^{-4}$ & \cr
& $200$ && $(0.03,~0.36)$ && $1.9\times 10^{-3}$ & \cr
& $210$ && $(0.10,~0.37)$ && $3.1\times 10^{-2}$ & \cr
& $220$ && $(0.16,~0.37)$ && $0.13$ & \cr
& $230$ && $(0.21,~0.36)$ && $0.32$ & \cr
& $240$ && $(0.26,~0.36)$ && $0.64$ & \cr
& $250$ && $(0.31,~0.36)$ && $1.07$ & \cr
& $260$ && $(0.35,~0.36)$ && $1.62$ & \cr
& $270$ && $(0.39,~0.36)$ && $2.28$ & \cr
height2pt&\omit&&\omit&&\omit&\cr
\noalign{\hrule}}}
\caption{The ``best values'' of the CKM parameters $(\rho,\eta)$ as a
function of the coupling constant $\fbd\protect\sqrt{\hat{B}_{B_d}}$,
obtained by a minimum $\chi^2$ fit to the experimental data, including the
renormalized value of $m_t=165 \pm 16$ GeV. We fix $\hat{B}_K=0.8$. The
resulting minimum $\chi^2$ values from the MINUIT fits are also given.}
\label{tabbk8}
\end{table}

\begin{table}
\hfil
\vbox{\offinterlineskip
\halign{&\vrule#&
   \strut\quad#\hfil\quad\cr
\noalign{\hrule}
height2pt&\omit&&\omit&&\omit&\cr
& $\fbd\sqrt{\hat{B}_{B_d}}$
 (MeV) && $(\rho,\eta)$ && $\chi^2_{min}$ & \cr
height2pt&\omit&&\omit&&\omit&\cr
\noalign{\hrule}
height2pt&\omit&&\omit&&\omit&\cr
& $110$ && $(-0.47,~0.16)$ && $3.40$ & \cr
& $120$ && $(-0.42,~0.20)$ && $1.98$ & \cr
& $130$ && $(-0.36,~0.24)$ && $1.09$ & \cr
& $140$ && $(-0.31,~0.28)$ && $0.58$ & \cr
& $150$ && $(-0.25,~0.32)$ && $0.31$ & \cr
& $160$ && $(-0.18,~0.35)$ && $0.18$ & \cr
& $170$ && $(-0.10,~0.38)$ && $0.15$ & \cr
& $180$ && $(-0.03,~0.40)$ && $0.18$ & \cr
& $190$ && $(0.05,~0.40)$ && $0.29$ & \cr
& $200$ && $(0.12,~0.41)$ && $0.50$ & \cr
& $210$ && $(0.19,~0.41)$ && $0.87$ & \cr
& $220$ && $(0.24,~0.40)$ && $1.33$ & \cr
& $230$ && $(0.30,~0.40)$ && $1.98$ & \cr
& $240$ && $(0.35,~0.40)$ && $2.79$ & \cr
height2pt&\omit&&\omit&&\omit&\cr
\noalign{\hrule}}}
\caption{The ``best values'' of the CKM parameters $(\rho,\eta)$ as a
function of the coupling constant $\fbd\protect\sqrt{\hat{B}_{B_d}}$,
obtained by a minimum $\chi^2$ fit to the experimental data, including the
renormalized value of $m_t=165 \pm 16$ GeV. We fix $\hat{B}_K=0.6$. The
resulting minimum $\chi^2$ values from the MINUIT fits are also given.}
\label{tabbk6}
\end{table}

\begin{table}
\hfil
\vbox{\offinterlineskip
\halign{&\vrule#&
   \strut\quad#\hfil\quad\cr
\noalign{\hrule}
height2pt&\omit&&\omit&&\omit&\cr
& $\fbd\sqrt{\hat{B}_{B_d}}$ (MeV)
 && $(\rho,\eta)$ && $\chi^2_{min}$ & \cr
height2pt&\omit&&\omit&&\omit&\cr
\noalign{\hrule}
height2pt&\omit&&\omit&&\omit&\cr
& $120$ && $(-0.38,~0.29)$ && $2.41$ & \cr
& $130$ && $(-0.31,~0.34)$ && $1.62$ & \cr
& $140$ && $(-0.23,~0.38)$ && $1.20$ & \cr
& $150$ && $(-0.14,~0.42)$ && $1.06$ & \cr
& $160$ && $(-0.05,~0.44)$ && $1.13$ & \cr
& $170$ && $(0.04,~0.45)$ && $1.41$ & \cr
& $180$ && $(0.12,~0.46)$ && $1.91$ & \cr
& $190$ && $(0.19,~0.46)$ && $2.65$ & \cr
height2pt&\omit&&\omit&&\omit&\cr
\noalign{\hrule}}}
\caption{The ``best values'' of the CKM parameters $(\rho,\eta)$ as a
function of the coupling constant $\fbd\protect\sqrt{\hat{B}_{B_d}}$,
obtained by a minimum $\chi^2$ fit to the experimental data, including the
renormalized value of $m_t=165 \pm 16$ GeV. We fix $\hat{B}_K=0.4$. The
resulting minimum $\chi^2$ values from the MINUIT fits are also given.}
\label{tabbk4}
\end{table}

We now discuss the ``combined fit" (Fit 2). Strictly speaking, this fit is
not on the same footing as the ``experimental fit" presented above, since
theoretical ``errors'' are not Gaussian. On the other hand, experimental
systematic errors are also not Gaussian, but it is common practice to treat
them as such, and to add them in quadrature with statistical errors. It is
in this spirit that we use this method. Since the coupling constants are
not known and the best we have are estimates given in the ranges in
eqs.~(\ref{BKrange1}) and (\ref{FBrange}), a reasonable profile of the
unitarity triangle at present can be obtained by letting the coupling
constants vary in these ranges. The resulting CKM triangle region is shown
in Fig.~\ref{rhoeta2}. As is clear from this figure, the allowed region is
enormous! We really know rather little about the unitarity triangle. Even
so, its allowed region is still reduced compared to the previous such
analyses, due to the knowledge of $\mt$. The preferred values obtained from
the ``combined fit" are
\beq
(\rho,\eta) = (-0.12,0.34) ~~~(\mbox{with}~\chi^2 = 1.1\times 10^{-3})~.
\eeq

For comparison, we also show the allowed region in the $(\rho,\eta)$ plane
for the case in which $\hat{B}_K = 0.6 \pm 0.2$ [eq.~(\ref{BKrange2})],
which is more favoured by chiral perturbation theory and QCD sum rules. The
CKM triangle region is shown in Fig.~\ref{rhoeta3}. Clearly, there is not
much difference between this figure and Fig.~\ref{rhoeta2}. The preferred
values obtained from this fit are
\beq
(\rho,\eta) = (-0.07,0.38) ~~~(\mbox{with}~\chi^2 = 0.13)~.
\eeq
We note that the preferred values for $(\rho,\eta)$ yield the following
preferred values for the matrix element ratio $\vert V_{td}/V_{ts} \vert$:
\begin{eqnarray}
\frac{\absvtd}{\absvts} &=& 0.26 ~~~[\mbox{for}~\hat{B}_K=0.6 \pm 0.2]~,
\nonumber \\
\frac{\absvtd}{\absvts} &=& 0.25 ~~~[\mbox{for}~\hat{B}_K=0.8 \pm 0.2]~.
\end{eqnarray}
Such values would make a number of measurements in CKM-suppressed rare
decays such as $B \to (\rho , \omega) + \gamma$ much more accessible
experimentally.

\section{$\xs$ and the Unitarity Triangle}

Mixing in the \bsbsbar\ system is quite similar to that in the \bdbdbar\
system. The \bsbsbar\ box diagram is again dominated by $t$-quark exchange,
and the mass difference between the mass eigenstates $\delms$ is given by a
formula analogous to that of eq.~(\ref{xd}):
\beq
\delms = \frac{G_F^2}{6\pi^2}M_W^2M_{B_s}\left(\fbbs\right)
\hat{\eta}_{B_s} y_t f_2(y_t) \vert V_{ts}^*V_{tb}\vert^2~.
\label{xs}
\eeq
Using the fact that $\vert V_{cb}\vert=\vert V_{ts}\vert$ (eq.~\ref{CKM}),
it is clear that one of the sides of the unitarity triangle, $\vert
V_{td}/\lambda V_{cb}\vert$, can be obtained from the ratio of $\delmd$ and
$\delms$,
\beq
\frac{\delms}{\delmd} =
 \frac{\hat{\eta}_{B_s}M_{B_s}\left(\fbbs\right)}
{\hat{\eta}_{B_d}M_{B_d}\left(\fbb\right)}
\left\vert \frac{V_{ts}}{V_{td}} \right\vert^2.
\label{xratio}
\eeq
All dependence on the $t$-quark mass drops out, leaving the square of the
ratio of CKM matrix elements, multiplied by a factor which reflects
$SU(3)_{\rm flavour}$ breaking effects. The only real uncertainty in this
factor is the ratio of hadronic matrix elements. Whether or not $\xs$ can
be used to help constrain the unitarity triangle will depend crucially on
the theoretical status of the ratio $\fbbs/\fbb$.

The lattice-QCD results for the hadronic quantities in the $B_s$ system are
\cite{Shigemitsu}
\begin{eqnarray}
\hat{B}_{B_s} \simeq \hat{B}_{B_d} &=& 1.2 \pm 0.2~, \nonumber \\
\frac{f_{B_s}}{f_{B_d}} &=& 1.16 \pm 0.1~,
\end{eqnarray}
while the correspondng values from QCD sum rules are \cite{Narison}
\begin{eqnarray}
\hat{B}_{B_s} \simeq \hat{B}_{B_d} &=& 1.0 \pm 0.15 ~, \nonumber \\
\frac{f_{B_s}}{f_{B_d}} &=& 1.16 \pm 0.05 ~.
\end{eqnarray}
In what follows, we will take $\xi_s \equiv (f_{B_s}
\sqrt{\hat{B}_{B_s}}) / (f_{B_d} \sqrt{\hat{B}_{B_d}}) = (1.16 \pm 0.1)$.
(The SU(3)-breaking factor in $\delms/\delmd$ is $\xi_s^2$.)

The mass and lifetime of the $B_s$ meson have now been measured at LEP and
Tevatron and their present values are $M_{B_s}=5368.0 \pm 3.7$ MeV and
$\tau(B_s)= 1.54^{+0.14}_{-0.13} \pm 0.05$ ps \cite{roudeau94}. We expect
the QCD correction factor $\hat{\eta}_{B_s}$ to be equal to its $B_d$
counterpart, i.e.\ $\hat{\eta}_{B_s} =0.55$. The main uncertainty in $\xs$
(or, equivalently, $\delms$) is now $\fbbs$. Using the determination of $A$
given previously, $\tau_{B_s}= 1.54 \pm 0.14$ (ps) and $\overline{\mt}=165
\pm 16$ GeV, we obtain
\begin{equation}
\xs = \left(19.4 \pm 6.9\right)\frac{\fbbs}{(230~\mbox{MeV})^2}~.
\end{equation}
The choice $f_{B_s}\sqrt{\hat{B}_{B_s}}= 230$ MeV corresponds to the
central value given by the lattice-QCD estimates, and with this our fits
give $\xs \simeq 20$ as the preferred value in the SM. Allowing the
coefficient to vary by $\pm 2\sigma$, and taking the central value for
$f_{B_s}\sqrt{\hat{B}_{B_s}}$, this gives
\begin{equation}
5.6 \leq \xs \leq 33.2~.
\label{bestxs}
\end{equation}
It is difficult to ascribe a confidence level to this range due to the
dependence on the unknown coupling constant factor. All one can say is that
the standard model predicts large values for $\xs$, most of which are above
the present experimental limits.

\begin{figure}
\vskip -2.0truein
\centerline{\epsfxsize 3.5 truein \epsfbox {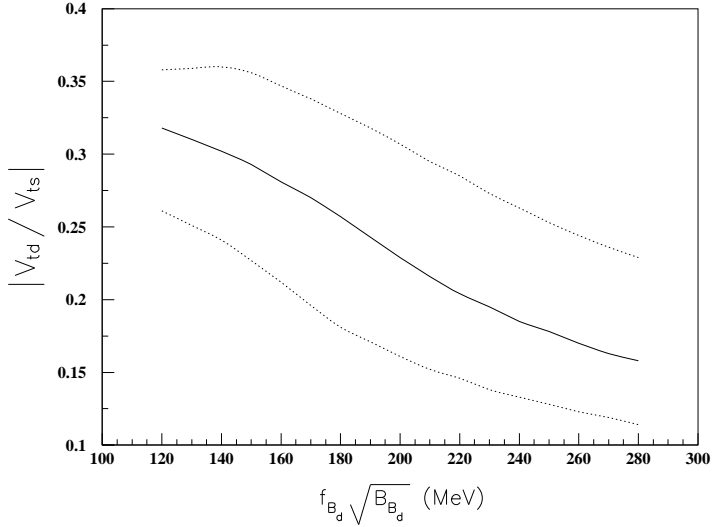}}
\vskip -1.0truein
\caption{Allowed values of the CKM matrix element ratio $\vert
V_{td}/V_{ts} \vert$ as a function of the coupling constant product
$f_{B_d}\protect\sqrt{\hat{B}_{B_d}}$, for $\hat{B}_K=0.8\pm 0.2$. The solid
line corresponds to the best fit values and the dotted curves correspond to
the maximum and minimum allowed values at 95 \% C.L.}
\label{vtdts}
\end{figure}

An alternative estimate of $\delms$ (or $\xs$) can also be obtained by
using the relation in eq.~(\ref{xratio}). Two quantities are required.
First, we need the CKM ratio $\vert V_{ts}/V_{td} \vert$. In
Fig.~\ref{vtdts} we show the allowed values (at 95\% C.L.) of the inverse
of this ratio as a function of $\fbd\sqrt{\hat{B}_{B_d}}$, for
$\hat{B}_K=0.8\pm 0.2$. From this one gets
\beq
2.76 \leq \left\vert {V_{ts} \over V_{td}} \right\vert \leq 8.4~.
\eeq
The second ingredient is the SU(3)-breaking factor which we take to be
$\xi_s = 1.16 \pm 0.1$, or $1.1 \le \xi_s^2 \le 1.6$.
  The result of the CKM fit can
 therefore be expressed as a $95\%$ C.L. range:
\beq
  10.3 \left(\frac{\xi_s}{1.16}\right)^2
      ~\leq ~\frac{\delms}{\delmd} ~\leq ~
  94.9 \left(\frac{\xi_s}{1.16}\right)^2 ~.
\eeq
Again, it is difficult to assign a true confidence level to $\delms/\delmd$
due to the dependence on $\xi_s$. The large allowed range reflects our poor
knowledge of the matrix element ratio $\vert V_{ts}/V_{td} \vert$, which
shows that this method is not particularly advantageous at present for the
determination of the range for $\delms$.

\begin{figure}
\vskip -2.2truein
\centerline{\epsfxsize 3.5 truein \epsfbox {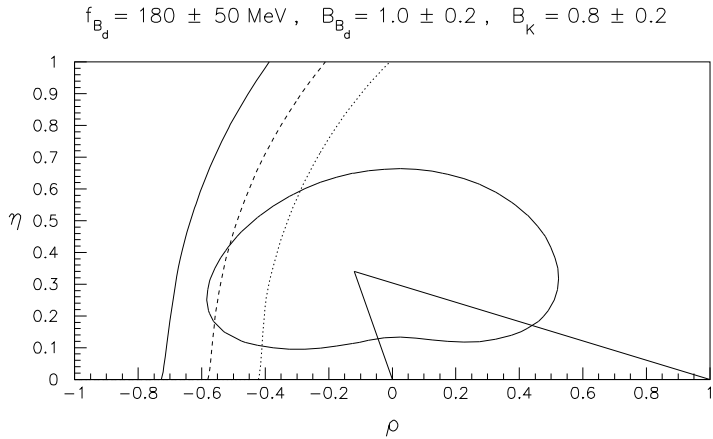}}
\vskip -1.4truein
\caption{Further constraints in $\rho$-$\eta$ space from the ALEPH bound
on $\delms$. The bounds are presented for 3 choices of the SU(3)-breaking
parameter: $\xi_s = 1.1$ (dotted line), $1.35$ (dashed line) and $1.6$
(solid line). In all cases, the region to the left of the curve is ruled
out.}
\label{xslimit}
\end{figure}

The ALEPH lower bound $\delms /\delmd > 11.3$ at $95\%$ C.L. can be turned
into a bound on the CKM parameter space $(\rho,\eta)$ by choosing a value
for the SU(3)-breaking parameter $\xi_s^2$. We assume three representative
values: $\xi_s^2 = 1.1$, $1.35$ and $1.6$, and display the resulting
constraints in Fig.~\ref{xslimit}. From this graph we see that the ALEPH
bound marginally  restricts the allowed $\rho$-$\eta$ region for small
values of $\xi_s^2$, but does not provide any useful bounds for larger
values.

Summarizing the discussion on $\xs$, we note that the lattice-QCD-inspired
estimate $f_{B_s} \sqrt{\hat{B}_{B_s}} \simeq 230$ MeV and the CKM fit
predict that $\xs$ lies between 6 and 33, with a central value around 20.
The upper and lower bounds and the central value scale as
$(f_{B_s}\sqrt{\hat{B}_{B_s}}/230 ~\mbox{MeV})^2$. The present constraints
from the lower bound on $\xs$ on the CKM parameters are marginal but this
would change with improved data. Of course, an actual measurement of $\xs$
would be very helpful in further constraining the CKM parameter space.

\section{CP Violation in the $B$ System}

It is expected that the $B$ system will exhibit large CP-violating effects,
characterized by nonzero values of the angles $\alpha$, $\beta$ and
$\gamma$ in the unitarity triangle (Fig.~\ref{triangle}) \cite{BCPasym}.
These angles can be measured via CP-violating asymmetries in hadronic $B$
decays. In the decays $\bdbarp \to \pi^+ \pi^-$, for example, one measures
the quantity $\sin 2\alpha$, and in $\bdbarp\to J/\psi K_S$, $\sin 2\beta$
is obtained. The CP asymmetry in the decay $\bsbarp\to D_s^\pm K^\mp$ is
slightly different, yielding $\sin^2 \gamma$.

These CP-violating asymmetries can be expressed straightforwardly in terms
of the CKM parameters $\rho$ and $\eta$. The 95\% C.L.\ constraints on
$\rho$ and $\eta$ found previously can be used to predict the ranges of
$\sin 2\alpha$, $\sin 2\beta$ and $\sin^2 \gamma$ allowed in the standard
model. The allowed ranges which correspond to each of the figures in
Fig.~\ref{rhoeta1}, obtained from Fit 1, are found in Table \ref{cpasym1}.
In this Table we have assumed that the angle $\beta$ is measured in
$\bdbarp\to J/\Psi K_S$, and have therefore included the extra minus sign
due to the CP of the final state.

\begin{table}
\hfil
\vbox{\offinterlineskip
\halign{&\vrule#&
   \strut\quad#\hfil\quad\cr
\noalign{\hrule}
height2pt&\omit&&\omit&&\omit&&\omit&\cr
& $\fbd\sqrt{\hat{B}_{B_d}}$ (MeV) && $\sin 2\alpha$ &&
$\sin 2\beta$ && $\sin^2 \gamma$ & \cr
height2pt&\omit&&\omit&&\omit&&\omit&\cr
\noalign{\hrule}
height2pt&\omit&&\omit&&\omit&&\omit&\cr
& $130$ && 0.36 -- 0.96 && 0.17 -- 0.41 && 0.08 -- 0.48 & \cr
& $155$ && 0.15 -- 1.0 && 0.26 -- 0.62 && 0.23 -- 1.0 & \cr
& $180$ && $-$1.0 -- 1.0 && 0.33 -- 0.81 && 0.37 -- 1.0 & \cr
& $205$ && $-$1.0 -- 1.0 && 0.40 -- 0.93 && 0.20 -- 1.0 & \cr
& $230$ && $-$1.0 -- 0.86 && 0.47 -- 0.99 && 0.15 -- 1.0 & \cr
height2pt&\omit&&\omit&&\omit&&\omit&\cr
\noalign{\hrule}}}
\caption{The allowed ranges for the CP asymmetries $\sin 2\alpha$, $\sin
2\beta$ and $\sin^2 \gamma$, corresponding to the constraints on $\rho$ and
$\eta$ shown in Fig.~\protect\ref{rhoeta1}. Values of the coupling constant
$\fbd\protect\sqrt{\hat{B}_{B_d}}$ are stated. We fix $\hat{B}_K=0.8$. The
range for $\sin 2\beta$ includes an additional minus sign due to the CP of
the final state $J/\Psi K_S$.}
\label{cpasym1}
\end{table}

\begin{figure}
\vskip -2.4truein
\centerline{\epsfxsize 2.6 truein \epsfbox {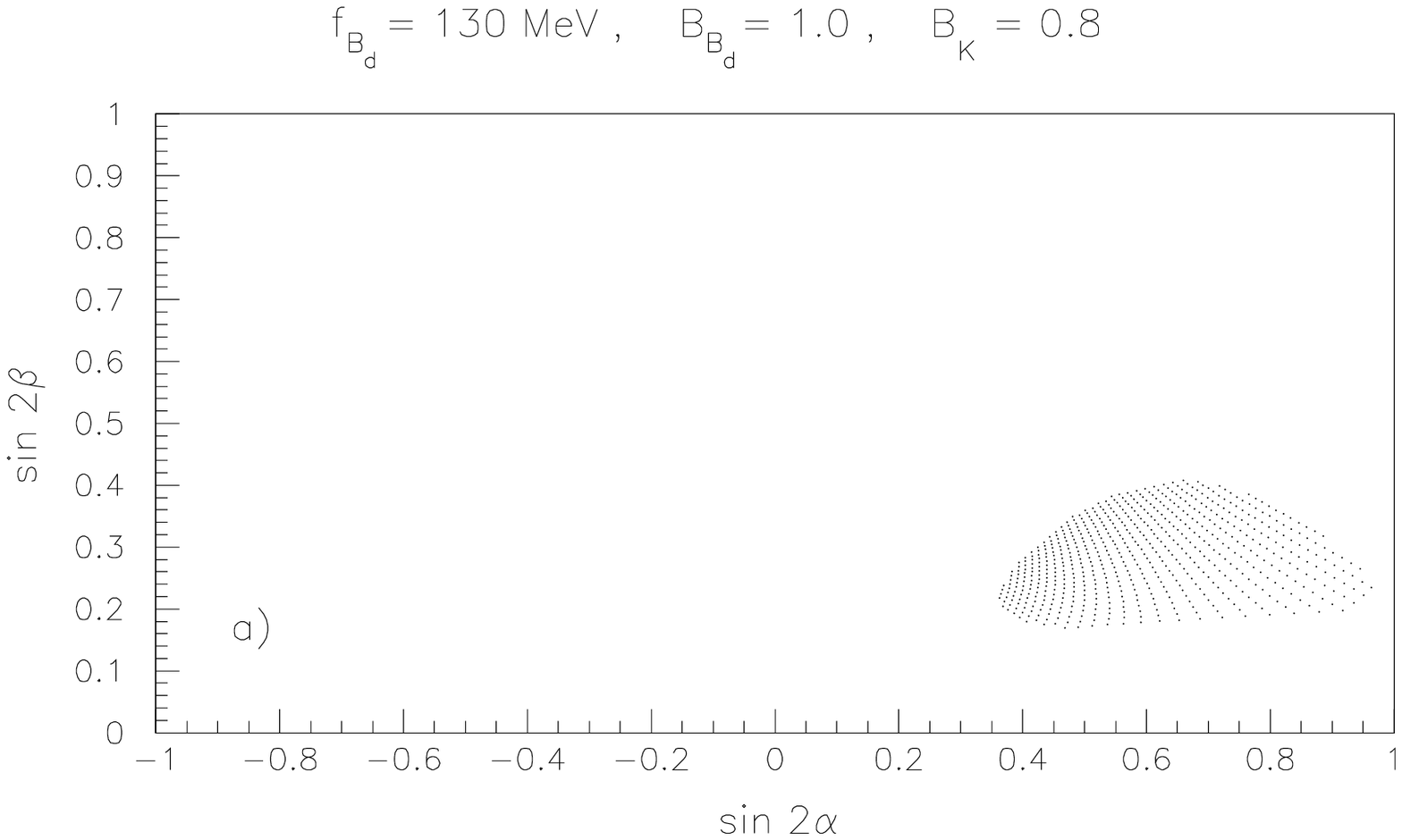}}
\vskip -1.8truein
\centerline{\epsfxsize 2.6 truein \epsfbox {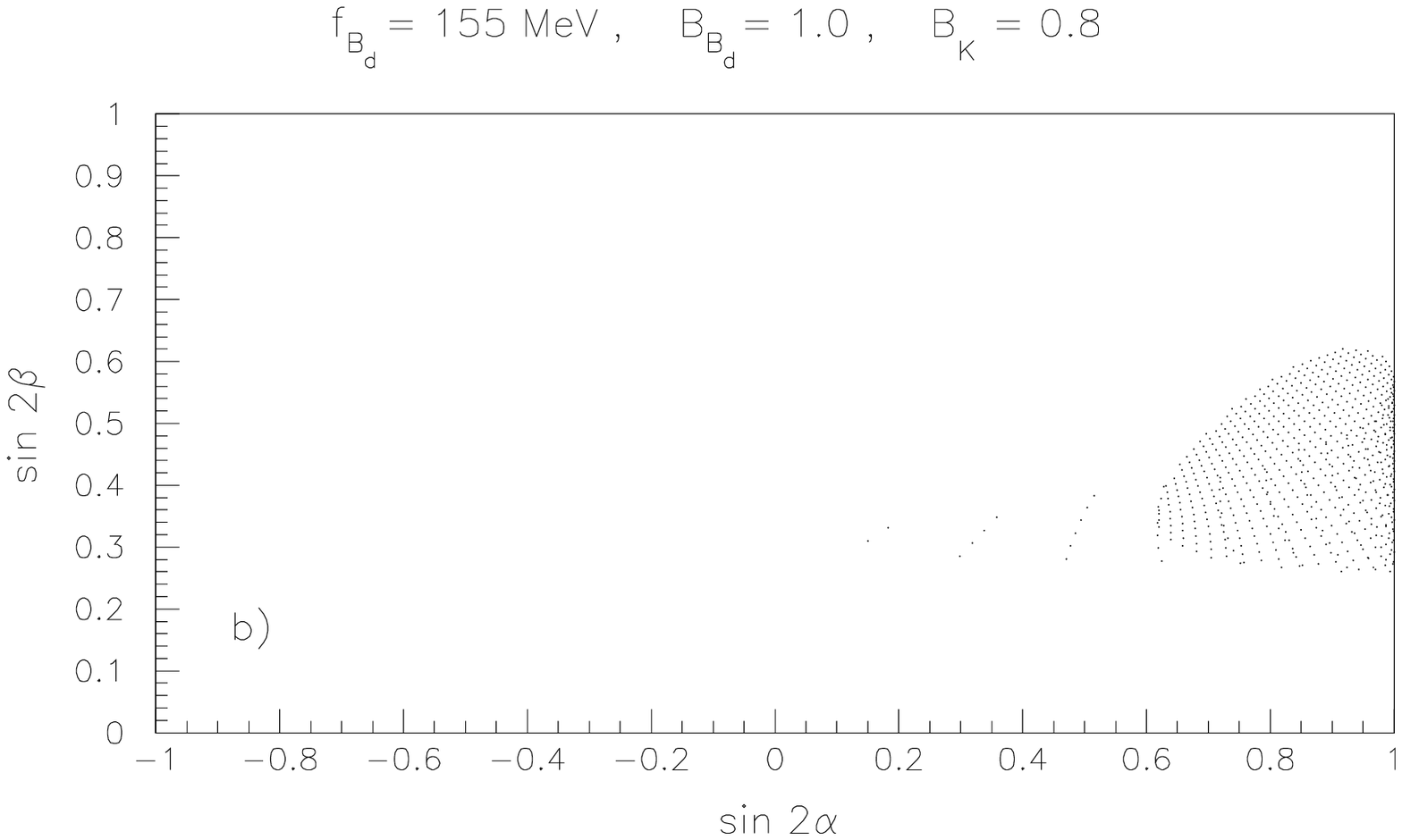}}
\vskip -1.8truein
\centerline{\epsfxsize 2.6 truein \epsfbox {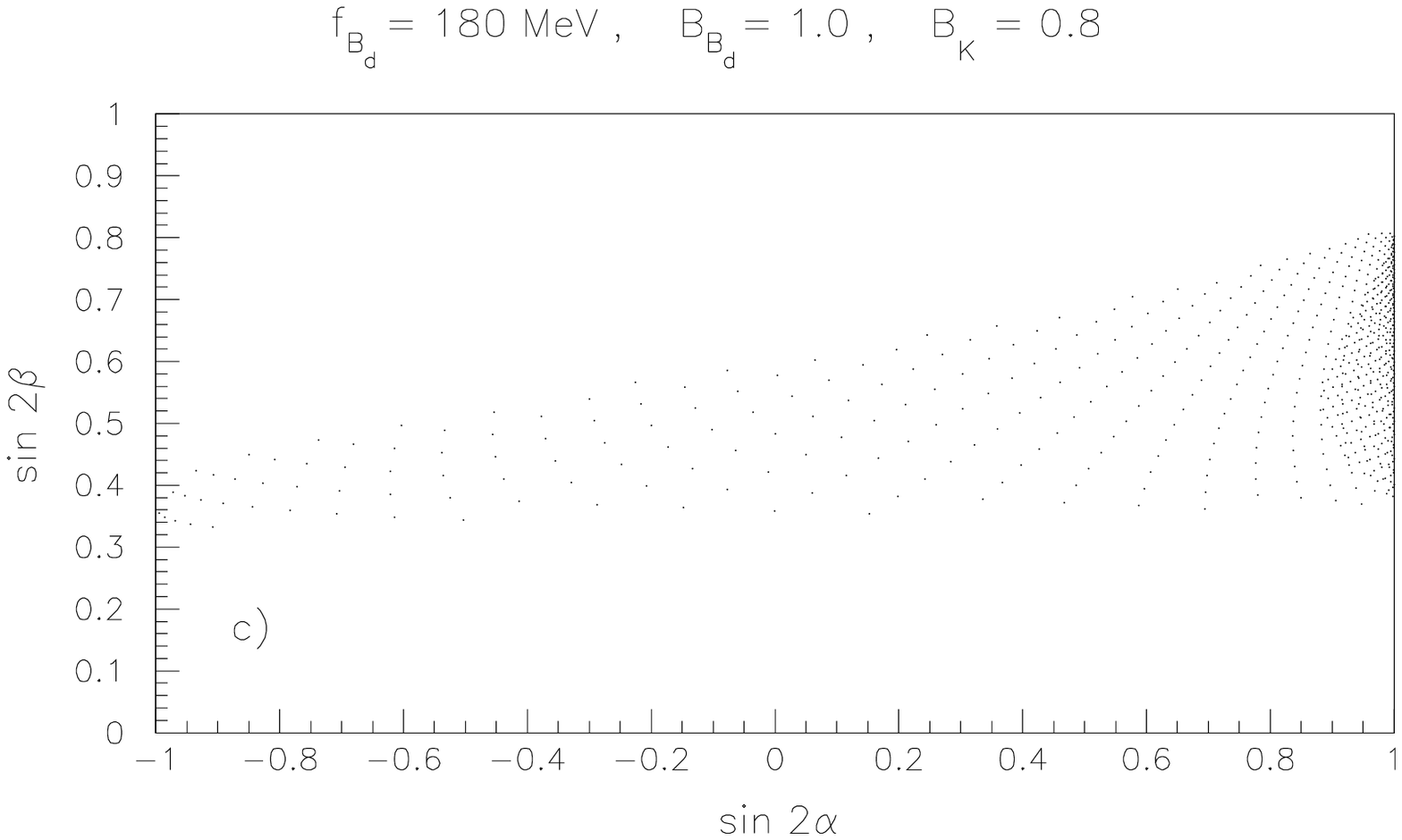}}
\vskip -1.8truein
\centerline{\epsfxsize 2.6 truein \epsfbox {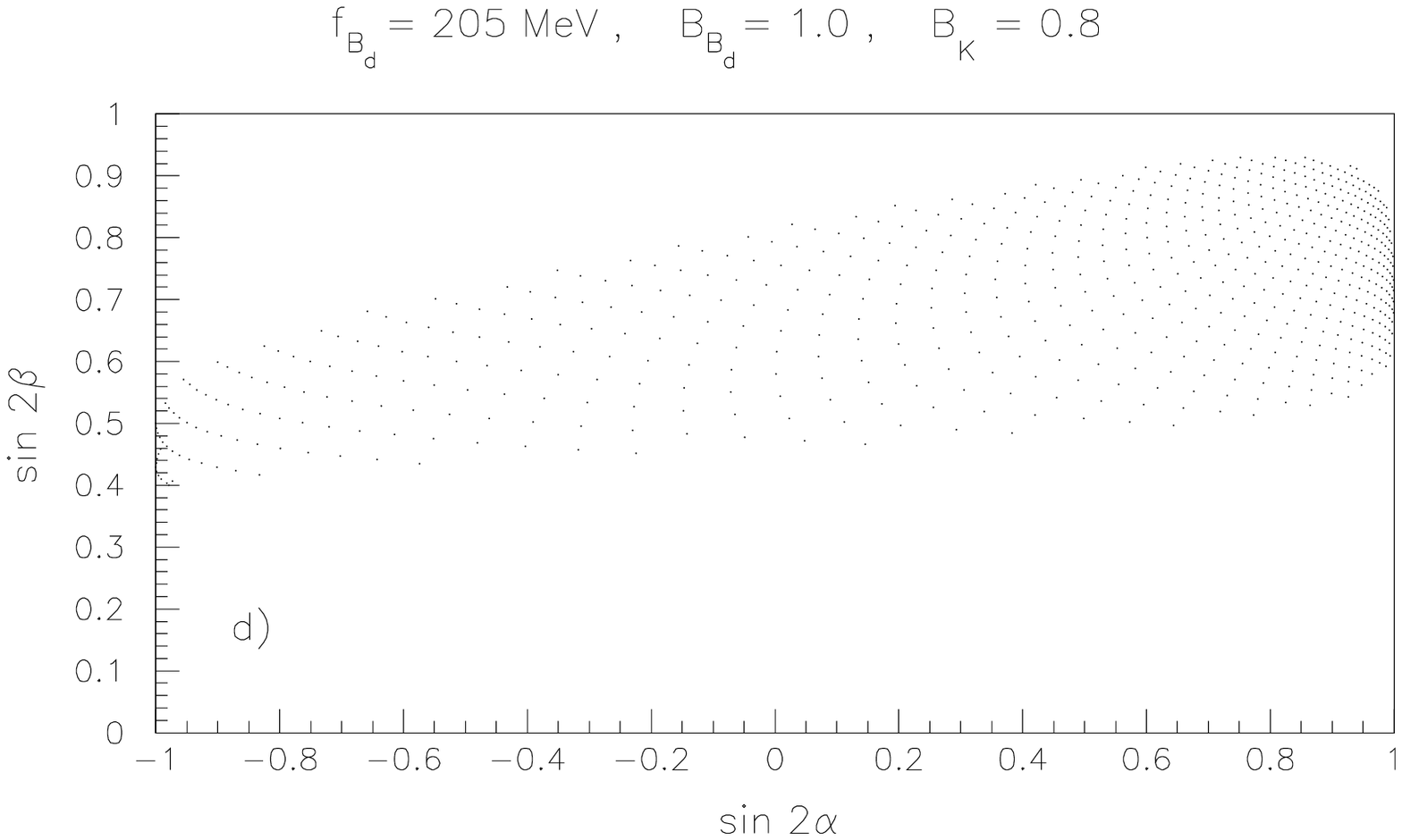}}
\vskip -1.8truein
\centerline{\epsfxsize 2.6 truein \epsfbox {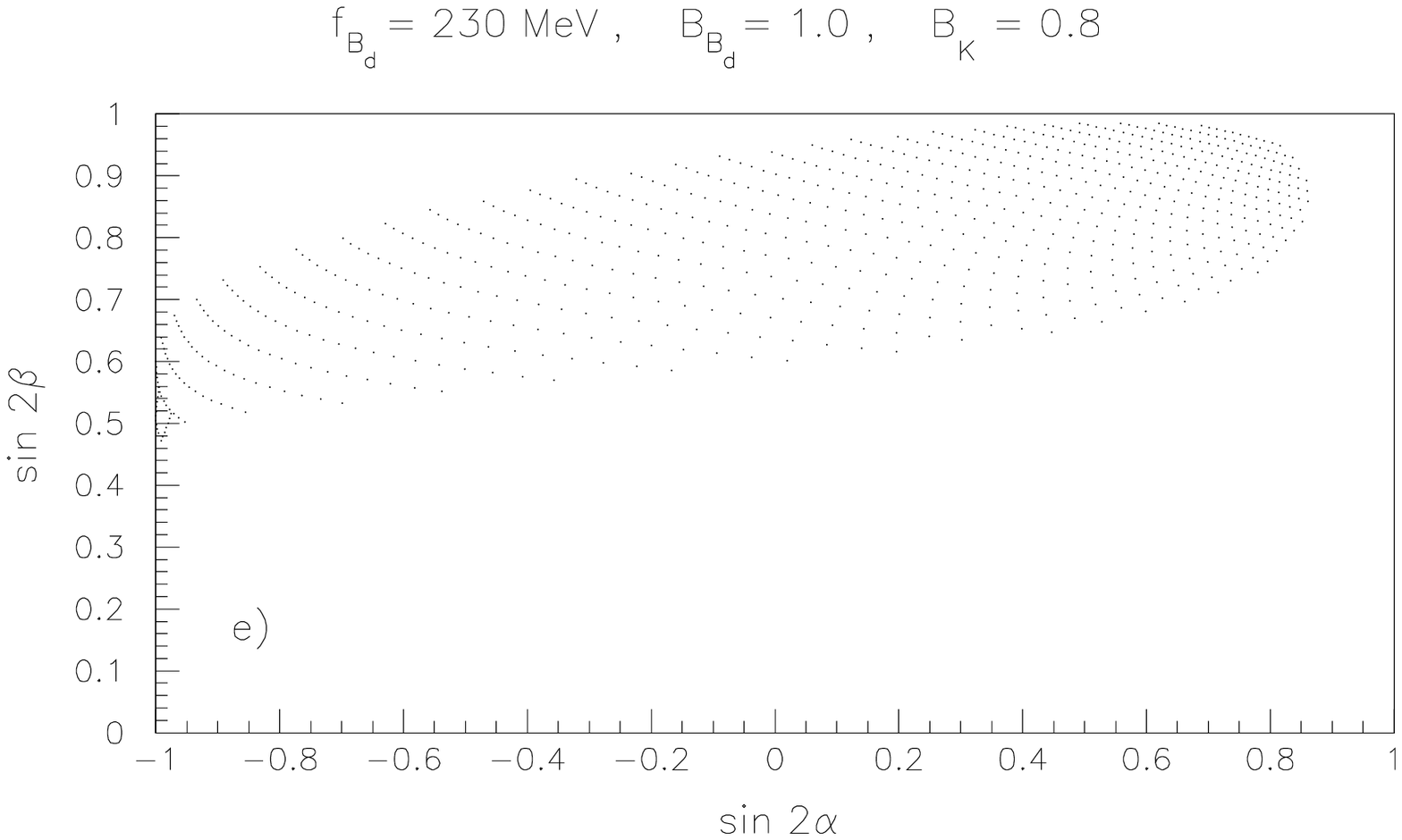}}
\vskip -1.0truein
\caption{Allowed region of the CP asymmetries $\sin 2\alpha$ and $\sin
2\beta$ resulting from the ``experimental fit" of the data for different
values of the coupling constant $\fbd\protect\sqrt{\hat{B}_{B_d}}$
indicated on the figures a) -- e). We fix $\hat{B}_K=0.8$.}
\label{alphabeta1}
\end{figure}

\begin{figure}
\vskip -2.2truein
\centerline{\epsfxsize 3.5 truein \epsfbox {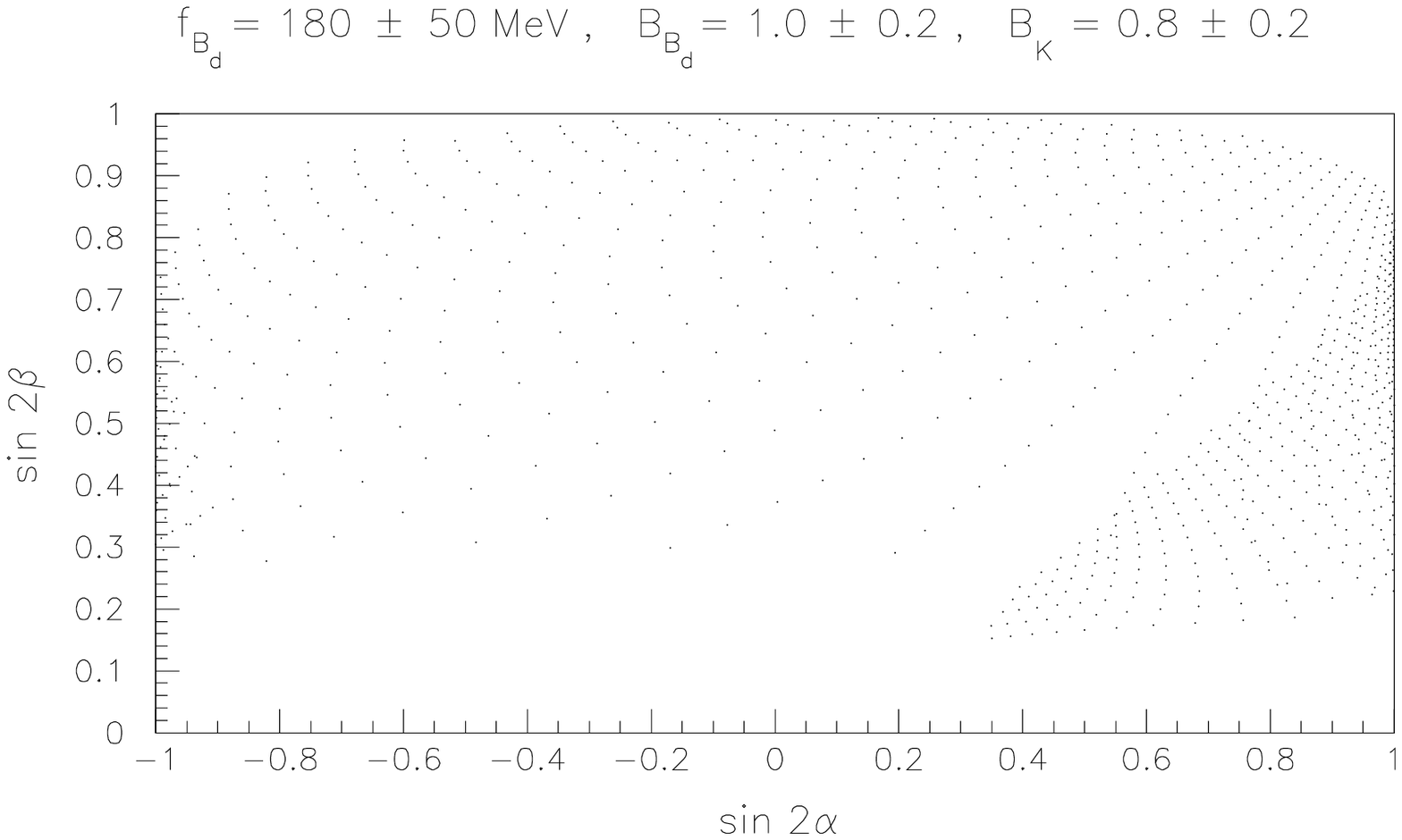}}
\vskip -1.4truein
\caption{Allowed region of the CP asymmetries $\sin 2\alpha$ and $\sin
2\beta$ resulting from the ``combined fit" of the data for the ranges for
$\fbd\protect\sqrt{\hat{B}_{B_d}} $ and $\hat{B}_K$ given in the text.}
\label{alphabeta2}
\end{figure}

Since the CP asymmetries all depend on $\rho$ and $\eta$, the ranges for
$\sin 2\alpha$, $\sin 2\beta$ and $\sin^2 \gamma$ shown in Table
\ref{cpasym1} are correlated. That is, not all values in the ranges are
allowed simultaneously. We illustrate this in Fig.~\ref{alphabeta1},
corresponding to the ``experimental fit" (Fit 1), by showing the region in
$\sin 2\alpha$-$\sin 2\beta$ space allowed by the data, for various values
of $\fbd\sqrt{\hat{B}_{B_d}}$. Given a value for
$\fbd\sqrt{\hat{B}_{B_d}}$, the CP asymmetries are fairly constrained.
However, since there is still considerable uncertainty in the values of the
coupling constants, a more reliable profile of the CP asymmetries at
present is given by our ``combined fit" (Fit 2), where we convolute the
present theoretical and experimental values in their allowed ranges. The
resulting correlation is shown in Fig.~\ref{alphabeta2}. From this figure
one sees that the smallest value of $\sin 2\beta$ occurs in a small region
of parameter space around $\sin 2\alpha\simeq 0.4$-0.6. Excluding this
small tail, one expects the CP-asymmetry in $\bdbarp\to J/\Psi K_S$ to be
at least 30\%.

\section{Summary and Outlook}

We summarize our results:

\smallskip

(i) We have presented an update of the CKM unitarity triangle following
from the additional experimental input of $\mt=174 \pm 16$ GeV
\cite{CDFmt}. The fits can be used to exclude extreme values of the
pseudoscalar coupling constants, with the range $110~\mbox{MeV} \leq
f_{B_d} \sqrt{\hat{B}_{B_d}} \leq 290~\mbox{MeV}$ still allowed for
$\hat{B}_K=1$. The lower limit of this range is quite
$\hat{B}_K$-independent, but the upper limit is strongly correlated with
the value chosen for $\hat{B}_K$. For example, for $\hat{B}_K=0.8$, $0.6$
and $0.4$, $f_{B_d} \sqrt{\hat{B}_{B_d}} \leq 270$, 230 and 180 MeV,
respectively, is required for a good fit. The solutions for $\hat{B}_K =
0.8 \pm 0.2$ are slightly favoured by the data as compared to the lower
values. These numbers are in very comfortable agreement with QCD-based
estimates from sum rules and lattice techniques. The statistical
significance of the fit is, however, not good enough to determine the
coupling constant more precisely. We note that the quality of fit for
$\hat{B}_K \leq 0.4$ is generally poor.

\smallskip

(ii) The allowed CKM unitarity triangle in the $(\rho,\eta)$-space is more
restricted than obtained previously without the top quark mass input.
However, the present uncertainties are still enormous -- despite the new,
more accurate experimental data, our knowledge of the unitarity triangle is
still poor. This underscores the importance of measuring CP-violating rate
asymmetries in the $B$ system. Such asymmetries are largely independent of
theoretical hadronic uncertainties, so that their measurement will allow us
to accurately pin down the parameters of the CKM matrix. Furthermore,
unless our knowledge of the pseudoscalar coupling constants improves
considerably, better measurements of such quantities as $\xd$ will not help
much in constraining the unitarity triangle. On this point, help may come
from the experimental front. It may be possible to measure the parameter
$\fbd$, using isospin symmetry, via the charged-current decay
$\bu\to\tau^\pm \nu_\tau$. With $\vert V_{ub}/V_{cb} \vert =0.08 \pm 0.03$
and $\fbd=180\pm 50~{\rm MeV}$, one gets a branching ratio
$BR(\bu\to\tau^\pm\nu_\tau)=(0.1$--$1.7)\times 10^{-4}$, with
$BR(\bu\to\tau^\pm\nu_\tau)=5.2\times 10^{-5}$ as the central value. This
lies in the range of the future LEP and asymmetric $B$-factory experiments,
though at LEP the rate $Z \to B_c X \to \tau^\pm \nu_\tau X$ could be just
as large as $Z \to B^\pm X \to \tau^\pm \nu_\tau X$. Along the same lines,
the prospects for measuring $(\fbd,\fbs)$ in the FCNC leptonic and photonic
decays of $\bd $ and $\bs$ hadrons, $(\bd,\bs)\to\mu^+\mu^-,
(\bd,\bs)\to\gamma\gamma$ in future $B$ physics facilities are not entirely
dismal \cite{ALIINT94}.

\smallskip

(iii) We have determined bounds on the ratio $\vert V_{td}/V_{ts} \vert$
from our fits. For $110~\mbox{MeV} \leq f_{B_d} \sqrt{\hat{B}_{B_d}} \leq
290~\mbox{MeV}$, i.e.\ in the entire allowed domain, at 95 \% C.L. we find
\beq
0.11 \leq \left\vert {V_{td} \over V_{ts}} \right\vert \leq 0.36~.
\eeq
The upper bound from our analysis is more restrictive than the current
experimental upper limit following from the CKM-suppressed radiative
penguin decays $BR(B \to \omega + \gamma )$ and $BR(B \to \rho + \gamma )$,
which at present yield at 90\% C.L. \cite{cleotdul}
\beq
\left\vert {V_{td} \over V_{ts}} \right\vert \leq 0.64 - 0.75~,
\eeq
depending on the model used for the SU(3)-breaking in the relevant
form factors \cite{SU3ff}.
Furthermore, the upper bound is now as good as that obtained from
unitarity, which gives $0.08 \leq \vert V_{td}/V_{ts} \vert \leq 0.36$
\cite{nir}, but the lower bound from our fit is more restrictive.

Note that the matrix element ratio $\vert V_{ub}/V_{cb} \vert$ is very
poorly determined. Our fits give:
\beq
     0.03 \leq  \frac{\absvub}{\absvcb} \leq 0.137 ~,
\label{utsides}
\eeq
It is important to reduce the present errors on this ratio. In particular,
a better determination of $\vert V_{ub}/V_{cb} \vert$ should be put as a
high priority item on the agenda of the ongoing experiments at CLEO and
LEP. We note here that the relations between the exclusive and inclusive
decays involving $b \to u \ell \nu_\ell$ and $b \to s \gamma$ transitions,
which have been discussed using the HQET methods, can be used to relate
$\absvub$ to $\absvts = \absvcb$.

\smallskip

(iv) Using the measured value of $\mt$, we find
\begin{equation}
\xs = \left(19.4 \pm 6.9\right)\frac{\fbbs}{(230 ~\mbox{MeV})^2}~.
\end{equation}
Taking $f_{B_s}\sqrt{\hat{B}_{B_s}}= 230$ (the central value of lattice-QCD
estimates), and allowing the coefficient to vary by $\pm 2\sigma$, this
gives
\begin{equation}
5.6 \leq \xs \leq 33.2~.
\end{equation}
No reliable confidence level can be assigned to this range -- all that one
can conclude is that the SM predicts large values for $\xs$, most of which
lie above the ALEPH 95\% C.L. lower limit of $\xs > 9.0$.

\smallskip

(v) The ranges for the CP-violating rate asymmetries parametrized by $\sin
2\alpha$, $\sin 2\beta$ and and $\sin^2 \gamma$ are determined at 95\% C.L.
to be
\begin{eqnarray}
&~& -1.0 \leq \sin 2\alpha \le 1.0~, \nonumber \\
&~& 0.17 \leq \sin 2\beta \le 0.99~, \\
&~& 0.08 \leq \sin^2 \gamma \le 1.0~. \nonumber
\end{eqnarray}
(For $\sin 2\alpha < 0.4$, we find $\sin 2\beta \ge 0.3$.)

\bigskip
\noindent
{\bf Acknowledgements}:
\bigskip

We thank David Cassel, Roger Forty, Peter Kim, Henning Schr\"oder, Vivek
Sharma and Ed Thorndike for providing updated analysis of $\absvcb$,
$B$-lifetimes, mixing parameters and rare $B$ decays, and for several
enlightening discussions. We thank Tony Pich for critical remarks on our
earlier manuscript. We thank Christoph Greub, Paul Langacker, Thomas
Mannel, Guido Martinelli, Yossi Nir, Rainer Sommer, and Nicolai Uraltsev
for discussions and helpful remarks. D.L. thanks Georges Azuelos for help
with the figures.



\end{document}